\documentclass{IEEEtran}

\usepackage{graphicx}
\usepackage[utf8x]{inputenc}
\usepackage[english]{babel}
\usepackage{amsmath}
\usepackage{amssymb}
\usepackage{color}
\usepackage{cite} 
\usepackage{amsthm}

\newtheorem*{lemma}{Lemma}
\newtheorem{thm}{Theorem}
\newcommand{\bx}{\boldsymbol{x}}
\newcommand{\by}{\boldsymbol{y}}
\newcommand{\bw}{\boldsymbol{w}}
\newcommand{\bH}{\boldsymbol{H}}
\newcommand{\bL}{\boldsymbol{\Sigma}}

\title{On the Use of Multiple Satellites to Improve the Spectral Efficiency of Broadcast Transmissions}
\author{Andrea Modenini, Alessandro Ugolini, Amina Piemontese,~\IEEEmembership{Member,~IEEE}, and Giulio Colavolpe,~\IEEEmembership{Senior Member,~IEEE}
\thanks{A. Modenini, A. Ugolini, A. Piemontese and G. Colavolpe are with Universit\`a di Parma, Dipartimento di Ingegneria dell'Informazione, Parco Area delle Scienze, 181A, I-43124 Parma, Italy, e-mail: andrea.modenini@unipr.it, alessandro.ugolini@unipr.it, amina.piemontese@unipr.it, giulio@unipr.it}
\thanks{This work is funded by the European Space Agency, ESA-ESTEC, Noordwijk, The Netherlands, under contract no. 4000109715/13/NL/FE. The view expressed herein can in no way be taken to reflect the official opinion of the European Space Agency.}
\thanks{The paper was presented in part at the 7th Advanced Satellite Mobile Systems Conference, 13th International Workshop on Signal Processing for Space Communications (ASMS/SPSC 2014), Livorno, Italy, September 2014.}
}

\begin{document}

\maketitle
\begin{abstract}
We consider the use of multiple co-located satellites to improve the spectral efficiency of broadcast transmissions. In particular, we assume that two satellites transmit on overlapping geographical coverage areas, with overlapping frequencies. We first describe the theoretical framework based on network information theory and, in particular, on the theory for multiple access channels. The application to different scenarios will be then considered, including the bandlimited additive white Gaussian noise channel with average power constraint and different models for the nonlinear satellite channel. The comparison with the adoption of frequency division multiplexing and with the Alamouti space-time block coding is also provided. The main conclusion is that a strategy based on overlapped signals is the most convenient in the case of no power unbalance, although it requires the adoption of a multiuser detection strategy at the receiver.
\end{abstract}


\section{Introduction}\label{sec:intro}
In today's satellite communication systems, the scarcity of frequency spectrum and the ever growing demand for data throughput has increased the need for resource sharing. In recent years, users of professional broadcast applications such as content contribution, distribution, and professional data services have demanded more spectrally efficient solutions.

Satellite service providers often have the availability of {\it co-located} satellites: two (or more) satellites are said to be co-located when, from a receiver on Earth, they appear to occupy the same orbital position. Co-location of satellites is typically used to cover the fully available spectrum by activating transponders on different satellites that cover non-overlapping frequencies or as a stand-by in-orbit redundancy, when the backup satellite is activated in case of failure of the main satellite. However, the second satellite can also be exploited to try to increase the capacity of the communications link. 

In this paper, we address a scenario in which the backup satellite is activated in addition to the main one, to improve the spectral efficiency (SE) of the overall communication system. The transmission from the two satellites can be coordinated, but through simple geometrical considerations it can be easily shown that even with a coverage area of a few tens of kilometers and two co-located signals separated in angle by a fraction of degree, time alignment is not possible. On the other hand, the considered system model can also represent a scenario where a single satellite with two transponders operating at the same frequency is employed and hence the two transmitted signals can be considered synchronous.

Here, we study the information rate (IR) achievable by a system where the two satellites transmit on overlapping geographical coverage areas with overlapping frequencies, and compare our results with that achievable by the frequency division multiplexing (FDM) strategy and with that achievable adopting the well-known Alamouti space-time block code~\cite{Al98}.

The two-satellites scenario has been studied in~\cite{ShChOt13,ChChOt13}, where the satellite channel is approximated as a linear additive white Gaussian noise (AWGN) channel, and the information theoretic analysis has been carried out under the limiting assumption of Gaussian inputs. We instead examine three different models for the system: the linear AWGN channel, the peak-power-limited AWGN channel~\cite{Sm71,Sh88,ShBa95}, and the satellite channel adopted in the 2nd-generation satellite digital video broadcasting (DVB-S2) standard~\cite{DVB-S2-TR}. The studied system is an instance of broadcast channel~\cite{CoTh06,Co75,CaSh03} with multiple transmitters. However, we are interested in a scenario in which the same information must be sent to every receiver. This situation corresponds, for example, to the delivery of a TV broadcast channel. We show that all these scenarios can be analyzed by means of network information theory and we will resort to multiple access channels (MACs)~\cite{ElCo80,CoTh06} with proper constraints.

Our analysis reveals that, if we allow multiuser detection, the strategy based on overlapping signals achieves higher SEs with respect to that achievable by using FDM. Interestingly, we show that there are cases in which a single satellite can outperform both these multiple satellites strategies, but not the Alamouti scheme.

The remainder of this paper is structured as follows: in Section~\ref{sec:sys} we present a general system model valid for all cases, and in Section~\ref{sec:mac} we briefly review the theory of MACs. In Sections~\ref{sec:fdm} and~\ref{sec:alam} we discuss the achievable rates by FDM and by the Alamouti space-time code. In Sections~\ref{sec:awgn_ap},~\ref{sec:awgn_pp}, and~\ref{sec:sat_ch} we analyze the three different scenarios and Section~\ref{sec:conclusion} concludes the paper.

\section{System Model}\label{sec:sys}

\begin{figure}
	\begin{center}
		\includegraphics[width=1.0\columnwidth]{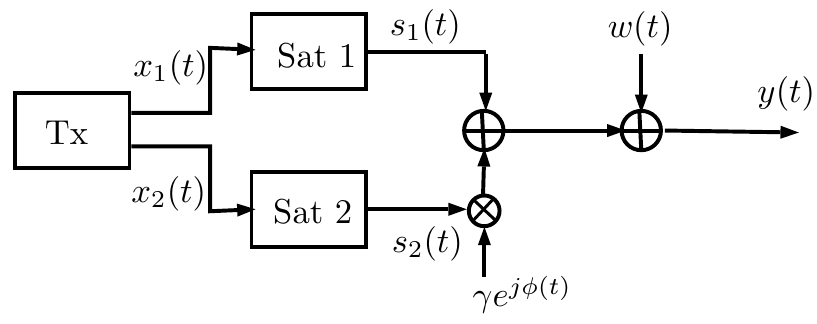}
		\caption{Block diagram of the analyzed system.}\label{fig:system}
	\end{center}
\end{figure}

Fig.~\ref{fig:system} depicts a schematic view of the baseband model we are considering. A single operator properly sends two separate data streams to the two satellites. The impact of the feeder uplink interference is considered negligible in this scenario. Data streams are linearly modulated 
signals, expressed as
\begin{equation}\label{eq:signals1}
	x_i(t) = \sum_k x_k^{(i)} p(t-kT) \quad i=1,2 \,,
\end{equation}
where $x_k^{(i)}$ is the $k$-th symbol transmitted on data stream $i$, $p(t)$ is the shaping pulse, and $T$ is the symbol time.

Each satellite then relays the signal, denoted as $s_i(t)$,\footnote{It can be $s_i(t)\neq x_i(t)$ due to the nonlinear transformation at the satellite transponder.} to several users scattered in its coverage area. For each user, the received signal is the sum of the two signals coming from the satellites, with a possible power unbalance $\gamma^2$ due to different path attenuations (we assume $1/2\le\gamma\le 1$). Without loss of generality, we assume that the attenuated signal is $s_2(t)$, otherwise we can exchange the roles of the two satellites. The received signal is also affected by a complex AWGN process $w(t)$ with power spectral density $N_0$. As mentioned, time alignment between the signals transmitted by the two satellites is not possible, since if the signals from the two satellites come perfectly aligned at a given receiver in the area, there will be other receivers for which a misalignment of a few symbols is observed. On the other hand, it is straightforward to show that our information-theoretic analysis does not depend on the time alignment of the two signals, and we will assume synchronous users to simplify the exposition. Hence, the received signal has the following expression
\begin{equation}\label{eq:rec}
	y(t)=s_1(t)+\gamma e^{j\phi(t)} s_2(t)+w(t)\,,
\end{equation}
where $s_1(t)$ and $s_2(t)$ are the signals at the output of the two satellites, and $\phi(t)$ is a possible phase noise process, caused by the instabilities of the oscillators. We assume that the phase noise is slow varying w.r.t. the signals' baud rate and perfectly known at the receiver. Signals $s_1(t)$ and $s_2(t)$ are transmitted with overlapping frequencies, and the overall signal has bandwidth~$W$. 

Since we are analyzing a broadcast scenario in which different receivers experience different (and unknown) levels of power unbalance, we impose that the two satellites transmit with the same rate. This constraint will be better clarified in the next sections. Channel state information is not available at the transmitter and no cooperation among the users is allowed. This is because the target is on broadcasting applications.

A simple alternative strategy to overlapping frequencies, that allows to avoid interference between the two transmitted signals, is FDM. The bandwidth $W$ is divided into two equal subbands assigned to the different satellites. An unequal subband allocation does not make sense since the power unbalance is different for different receivers in the coverage area and, in any case, unknown to the transmitter. In this case, the received signal has expression
\begin{equation}\label{eq:fdm}
 	y(t)= s_1(t)e^{j\pi f_c t} + \gamma e^{-j\pi f_ct + j\phi(t)} s_2(t) + w(t)\,,
\end{equation}
where $f_c$ is the frequency separation between the two signals.

Another possible alternative to avoid interference between the two signals is the use of the Alamouti space-time block code~\cite{Al98}, consisting in the two satellites exchanging the transmitted signals in two consecutive transmissions. Unlike the two previous strategies, its  classical implementation requires a perfect alignment in time of the signals received from the two satellites. However, in Appendix~\ref{app:alamouti} we  will describe an alternative implementation working in the presence of a delay which can be different for different receivers.

In the following, these transmission strategies will be compared by using the overall SE of the system as a figure of merit. The SE is defined as
$$
{\rm SE}=\frac{I}{TW}\quad {\rm [bit/s/Hz]},
$$
where $I$ is the maximum mutual information of the channel. However, since in the scenario of interest the values of $T$ and $W$ are fixed, without loss of generality we will assume that $TW=1$ and we will refer to the terms IR and SE interchangeably.

\section{Multiple Access Channels}\label{sec:mac}
In this section, we review some results on classical MACs~\cite{CoTh06}. We consider the transmission of independent signals from the two satellites.\footnote{We will explain later in Section~\ref{sec:awgn_ap} why this is the best choice for the signals transmitted from the two satellites.} We denote by $R_1$ the SE of the first satellite and by $R_2$ that of the second satellite. At this point, we make no assumptions on the channel inputs, since a better characterization of the input distributions is presented in the next sections. However, independently of the form assumed by the input distribution, the boundaries of the SE region can be expressed, for each fixed signal-to-noise ratio (SNR), as~\cite{CoTh06}
\begin{align}
	R_1 & \le I(x_1;y|x_2) \triangleq I_1\nonumber \\
	R_2 & \le I(x_2;y|x_1) \triangleq I_2\nonumber \\
	R_1+R_2 & \le I(x_1,x_2;y) \triangleq I_{\rm J}\nonumber\,,
\end{align}
where $I(x_1;y|x_2)$, $I(x_2;y|x_1)$ and $I(x_1,x_2;y)$  represent, respectively, the mutual information between $x_1$ and $y$ conditioned to $x_2$, that between $x_2$ and $y$ conditioned to $x_1$ and that between the couple ($x_1,x_2$) and $y$; we have omitted the dependence on $t$ and we adopt definitions $I_1$, $I_2$, and $I_{\rm J}$  to simplify the notation.

\begin{figure}
	\centering
	\includegraphics[width=0.8\columnwidth]{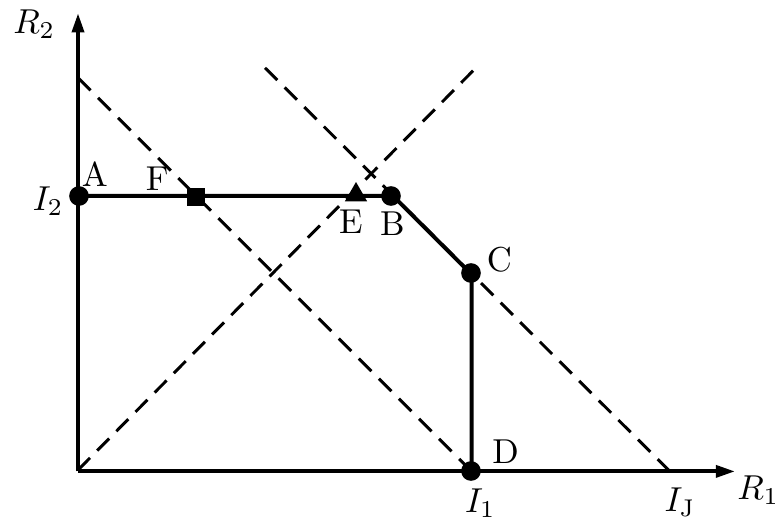}
	\caption{Achievable rate region in the case $I_2<I_1$. Point D is the maximum achievable SE from satellite 1 to the receiver when satellite 2 is not sending information. Point C is the maximum rate at which satellite 2 can transmit as long as satellite 1 transmits at its maximum rate. The maximum of the sum of the SEs is obtained on points of the segment B-C, which can be achieved by joint decoding.}
	\label{fig:cap_reg_ex}
\end{figure}

Fig.~\ref{fig:cap_reg_ex} is useful to gain a better understanding of the behavior of SE regions. Point D corresponds to the maximum achievable SE from satellite 1  to the receiver when satellite 2  is not sending any information. Point C corresponds to the maximum rate at which satellite 2  can transmit as long as satellite 1  transmits at its maximum rate.\footnote{If we exchange the role of the two satellites, the same considerations hold for points A and B instead of D and C.} The maximum of the sum of the SEs, however, is obtained on points of the segment B-C; these points can be achieved by joint decoding of both signals. It does not make sense to adopt different rates for the two satellites, since each satellite ignores whether its signal will be attenuated or not and this attenuation will vary for different receivers. As a consequence, the only boundary point of the SE region we can achieve is point E, which lies on the line $R_1=R_2$.  We define $I_{\rm J,p}$ the pragmatic sum of rates corresponding to point E: it is easy to see, through graphical considerations, that 
\begin{equation}
	I_{\rm J,p}= \min ( I_{\rm J}, 2I_2 )\nonumber \,.
\end{equation}
Point F is the intersection between the capacity region and the line $R_2=-R_1+I_1$ and it corresponds to a sum-rate equal to $I_1$. The position of point E depends on the power and on the power unbalance. Depending on these two values, E can be found in different positions: in particular, if E lies on the left of F, we can notice that $I_{\rm J,p}<I_1$, hence it is convenient to use a single satellite with rate $I_1$ rather than activating the second satellite.

\section{Achievable Rates by Frequency Division Multiplexing}\label{sec:fdm}
Since the two signals transmitted by the FDM model~\eqref{eq:fdm} operate on disjoint bandwidths, they are independent and the IR achievable by this system is equal, in case $\gamma=1$, to that of a single transmitter with double SNR. We define by $I_{\rm FDM}$ the achievable rate by FDM, and by $I_{\rm FDM,p}$ that achievable by FDM under the equal rate constraint. The latter is clearly equal to twice the minimum SE of the two subchannels. We demonstrate that the rate achievable by FDM is always lower or equal than that achievable with two signals with overlapping frequencies in the absence of nonlinear distortions; the same result holds for the pragmatic rates and can be stated in the following theorem, whose proof can be found in Appendix~\ref{app:fdm}.
\begin{thm}\label{thm:fdm}
	Let us consider the ideal multiple access channel 
	\begin{equation}
		y(t)  =  x_1(t) + \gamma x_2(t) + w(t)  \label{eq:chJ} \,.
	\end{equation}
	The following inequalities hold
	\begin{eqnarray}
		I_{\mathrm{J}} & \geq & I_{\rm FDM} \label{eq:i1} \\ 
		I_{\mathrm{J,p}} & \geq & I_{\rm FDM,p}  \label{eq:i3} 
	\end{eqnarray}
	with equality if and only if $x_1(t)$ and $x_2(t)$ are Gaussian random processes and $\gamma^2=0$ dB.
\end{thm}

The theorem, beyond the mathematical proof, has a practical explanation. The use of a second satellite, besides increasing the overall transmitted power, makes the distribution of $x_1(t)+\gamma x_2(t)$ closer to a Gaussian distribution (see the Berry-Esséen theorem~\cite{Fe71}). Thus, a sort of shaping gain must be added to the gain arising from the higher power.

The strategy based on FDM is perfectly equivalent, in terms of SE, to a strategy based on time-division multiplexing (TDM), in which time is divided into slots of equal length, and each satellite is allotted a slot during which only that satellite transmits and the other remains silent. During its slot, each satellite is allowed to use twice the power. However, on satellites, due to peak power constraints, it is not possible to double the power and the satellite amplifiers are not conceived for power bursts. Hence TDM strategy will not be considered.

\section{Achievable Rates by the Alamouti Scheme}\label{sec:alam}
We now consider the application of the Alamouti scheme~\cite{Al98}. The two satellites first transmit $x_1(t)$ and $x_2(t)$ and then $x^*_2(t)$ and 
$-x^*_1(t)$, respectively. The rate $I_{\rm A}$, achievable by the Alamouti scheme, satisfies the following theorem, proved in Appendix~\ref{app:alam}.

\begin{thm}\label{thm:alam}
		Let us consider the ideal multiple access channel 
		$$
			y(t)  =  x_1(t) + \gamma x_2(t) + w(t)  \,,
		$$
		where $x_i(t)$, $i=1,2$ are random processes such that $-x_i(t)$ has the same finite-dimensional distributions as $x_i(t)$.
		The following inequality holds
		$$		
			I_{\mathrm{J}} \underset{(a)}{\geq}  I_{\rm A}  \underset{(b)}{\geq}   I_{\rm FDM} 
		$$
		with equality in $(a)$ if and only if  $x_1(t)$ and $x_2(t)$ 	are independent Gaussian random processes with the same variance, and in $(b)$ if and only if $\gamma^2=0$ dB.
	\end{thm}

Theorem \ref{thm:alam} shows that the Alamouti scheme IR is between the ones achievable by two overlapping signals and by FDM. However, it has the interesting feature that it is not degraded by the equal rates constraint, being $I_{\rm A, p}=I_{\rm A}$, where the subscript p stands for pragmatic. This is due to the fact that both signals are transmitted once by the satellite with no attenuation and once by the satellite with attenuation $\gamma$. Hence, while it is always true that $I_{\rm A}\geq I_{\rm FDM,p}$, it can happen that $I_{\rm A}\geq I_{\rm J,p}$.

\section{Additive White Gaussian Noise Channel with Average Power Constraint}\label{sec:awgn_ap}
A first case study, useful to draw some preliminary considerations about the theoretical limits for the system under consideration, is the classical AWGN channel with average power constraint. For this case, we have that the two satellites of Fig.~\ref{fig:system} have no effect on the 
signal, hence the received signal reads
\begin{equation}
	y(t)=x_1(t)+\gamma e^{j\phi(t)} x_2(t)+w(t)\nonumber\,.
\end{equation}
We express the average power constraint as
\begin{equation}
	{\rm E}\left[|x_i(t)|^2\right] \le P \quad i=1,2 \nonumber\,,
\end{equation}
where $P$ is the maximum allowed average power. 

For this channel, the capacity is reached with independent Gaussian inputs, $p(t)=\mathrm{sinc}(t/T)$, and $TW=1$~\cite{Sh48}. A sufficient statistic is derived by sampling the output of a low pass filter \cite{Sh48,MeOePo94}. Since we are assuming slow-varying phase noise, the observable is
$$
	y_k = x_k^{(1)} + \gamma e^{j\phi_k} x_k^{(2)} + w_k\,,
$$
where $\phi_k=\phi(kT)$. The phase noise does not change the statistics, and hence the SE $I_{\rm J}$ is given by the classical Shannon capacity, taking into account the total transmitted power, and reads
\begin{equation}
	I_{\rm J}=\log_2\left(1+(1+\gamma^2)\frac{P}{N}\right) \nonumber\,,
\end{equation}
where $N=N_0W$ is the noise power in the considered bandwidth. If, instead, we adopt the FDM model~\eqref{eq:fdm}, the SE can be computed as the average of the SEs of two subchannels, each transmitting on half the bandwidth:
\begin{equation}
	I_{\rm FDM}=\frac{1}{2}\log_2\left(1+2\frac{P}{N}\right)+ \frac{1}{2}\log_2\left(1+2\gamma^2\frac{P}{N}\right) \nonumber\,.
\end{equation}
When we introduce the equal rate constraint, it is straightforward to show that we have the following pragmatic SEs
\begin{eqnarray}
 I_{\rm J,p} & = & \min\left(I_{\rm J},2\log_2\left( 1+\gamma^2\frac{P}{N} \right) \right)\nonumber\, \\
 I_{\rm FDM,p}& = & \log_2\left(1+2\gamma^2\frac{P}{N}\right)    \,. \nonumber
\end{eqnarray}

\begin{figure}
	\centering
	\includegraphics[width=1.0\columnwidth]{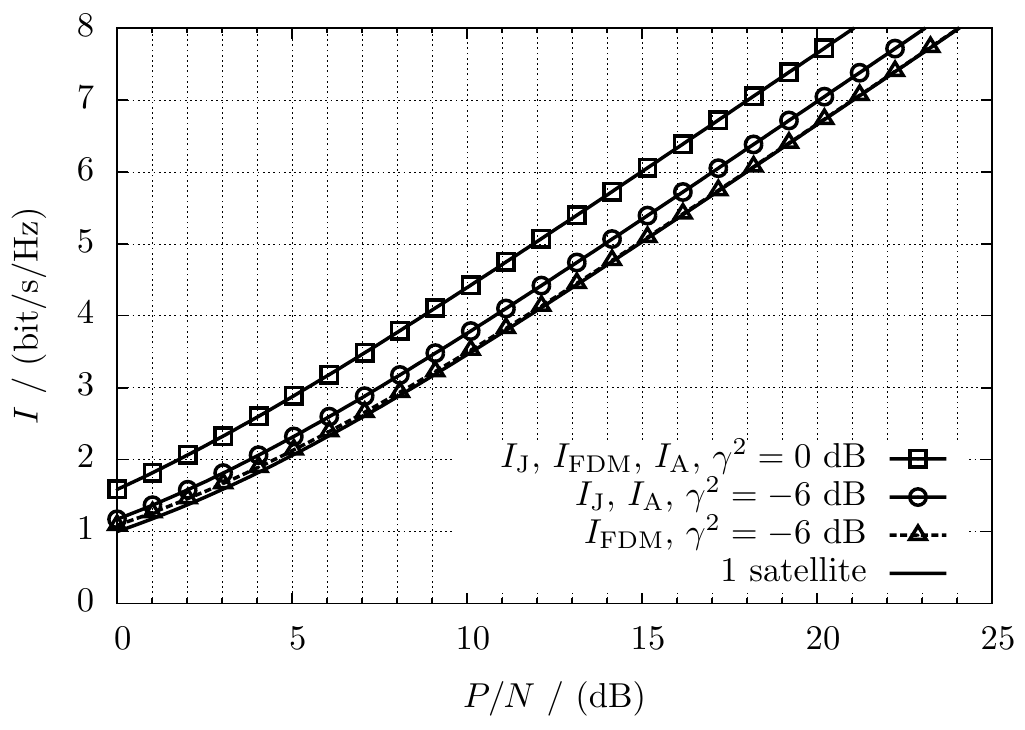}
	\caption{Joint spectral efficiency for different values of $\gamma$ (AWGN channel with average power constraint).}
	\label{fig:cap_joint_apc}
\end{figure}

In Fig.~\ref{fig:cap_joint_apc} we show the SE $I_{\rm J}$ as a function of $P/N$, for different values of power unbalance $\gamma$, together with the SE that can be achieved when a single satellite is available (\mbox{$\gamma\rightarrow 0$}). In this case, the performance of the Alamouti scheme is exactly the same as $I_{\rm J}$, as foreseen by Theorem~\ref{thm:alam}. The figure also shows $I_{\rm FDM}$ for the same values of~$\gamma$. We see that FDM is capacity-achieving when $\gamma=1$ (i.e., $I_{\rm J}=I_{\rm FDM}$ when $\gamma=1$, as also clear from the equations and as foreseen by Theorem~\ref{thm:fdm}) but it is suboptimal in the case of power unbalance. 

\begin{figure}
	\centering
	\includegraphics[width=1.0\columnwidth]{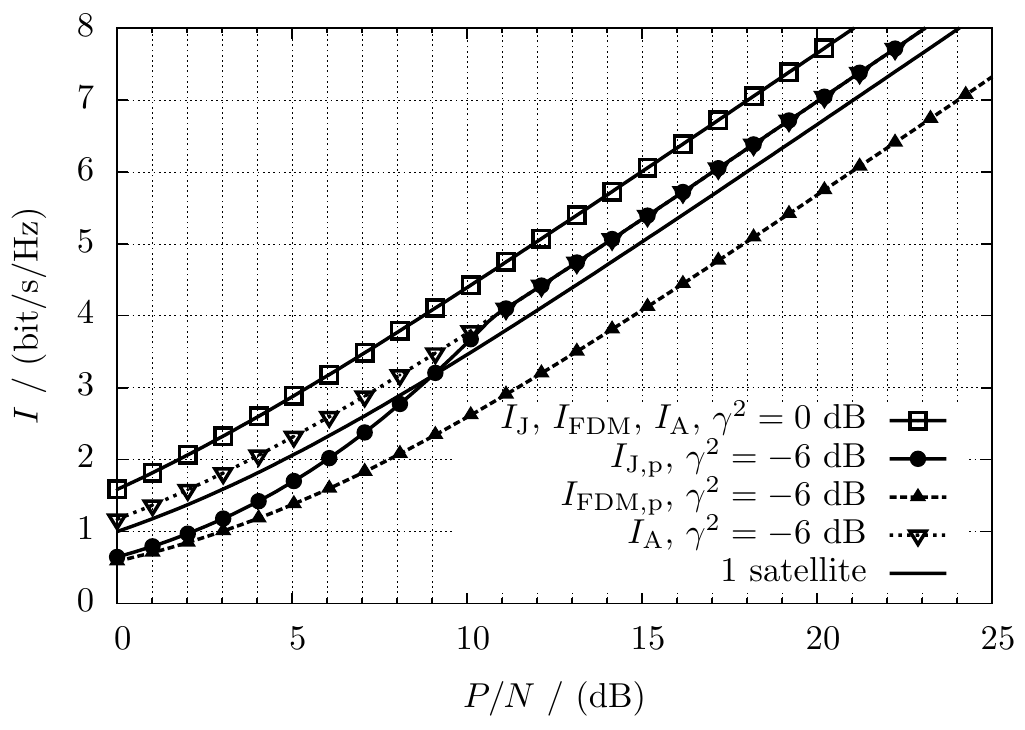}
	\caption{Pragmatic spectral efficiency for different values of $\gamma$ (AWGN channel with average power constraint).}
	\label{fig:cap_eq_rate_apc}
\end{figure}

In Fig.~\ref{fig:cap_eq_rate_apc} we report the pragmatic SEs for the cases of Fig.~\ref{fig:cap_joint_apc}. For signals with overlapping frequencies, with power unbalance $\gamma^2\neq 0$ dB, $I_{\rm J,p}$ is lower than $I_{\rm J}$ only in the range of low $P/N$ values, corresponding to the case $2I_2 < I_{\rm J}$. The transition is indicated by the change of slope in the curve. We also see that, for high power unbalance, a portion of $I_{\rm J,p}$ lies below single-satellite SE.

In case of FDM, we clearly see how the user with the lower SE limits $I_{\rm FDM,p}$. The curves coincide for $\gamma=1$, while they suffer from a significant performance loss w.r.t. $I_{\rm FDM}$ for high values of power unbalance. If the power unbalance is very high, FDM performs even worse than a single satellite. Finally we can notice that, when $\gamma^2\neq 0$ dB, $I_{\rm A}>I_{\rm J,p}$ for low $P/N$ values.

At the end of this section, we would like to motivate our choice of transmitting independent signals from the two satellites. Let us consider the opposite scenario where the same signal is transmitted from the two satellites. The received signal $y(t)$ can thus be expressed as
\begin{align}
y(t)&=x(t)+\gamma e^{j \phi(t)}x(t-\tau)+w(t) \nonumber
\end{align}
where $x(t)$ is the transmitted signal, and $\tau$ the difference between the propagation delays of the two satellites. The received sample at time $kT$ reads
$$
	y_k  =  x_k + \gamma e^{j\phi_k} \sum_i \mathrm{sinc}(i-\tau/T) x_{k-i} + w_k \,.
$$ 
The channel is thus equivalent to a time-varying frequency-selective channel 
\begin{equation}
	 y_k= \sum_{i} h_{k,i} x_{k-i} + w_k \label{eq:fs_ch}
\end{equation}
with impulse response $h_{k,i}=\gamma e^{j\phi_k} \mathrm{sinc}(i-\tau/T)$ for $i\neq 0$ and $h_{k,0}=1+ \gamma e^{j\phi_k}\mathrm{sinc}(-\tau/T)$. 
As already said, $\phi_k$ is assumed slowly varying with respect to the symbol interval but, due to the oscillators' instabilities, it will be assumed with a coherence time shorter than the codeword length. Hence, we are interested in the ergodic rate obtained by averaging the information rate that can be obtained with a given value of $\phi$. Independently of the value of $\tau$, the average signal power is 
\begin{equation}
\mathrm{E}\left[ \left|\sum_i h_{k,i}x_{k-i}\right|^2 \right]= (1+\gamma^2)P \label{eq:SNR}
\end{equation}
and it can be show that the ergodic rate cannot be higher than $\log_2\left( 1 + (1+\gamma^2) \frac{P}{N} \right)$, the rate achievable with independent signals (see Appendix~\ref{app:ergodic} for a detailed proof).


\section{Additive White Gaussian Noise Channel with Peak Power Constraint}\label{sec:awgn_pp}
As a first step to the theoretical characterization of our satellite transmission problem, we consider the case a peak-power-limited signal rather than an average-power-limited one. The adoption of a peak power constraint comes naturally from the use of a saturated nonlinear high-power amplifier (HPA) at the satellite. However, there is no expression for the channel capacity in this scenario, but only bounds are available~\cite{Sh88}. For this reason, we concentrate on the study of a simplified discrete-time channel, where the peak power constraint is imposed on information symbols~\cite{ShBa95}.

In this section, we repeat the analysis of Section~\ref{sec:awgn_ap} in a peak-power-limited scenario. We first review the results in~\cite{ShBa95} for the case of a single transmitter, then we extend the reasoning to the case of two transmitters.

\subsection{Analysis for Single Transmitter}\label{sec:1tx}
If we assume that $\gamma\rightarrow0$, the model \eqref{eq:rec} simplifies to the following discrete-time memoryless channel model
\begin{equation}\label{eq:model_pp}
	y_k=x_k+w_k \,,
\end{equation}
where $y_k$ is the observable, $x_k= x_k^{(1)}$ is the $k$-th symbol transmitted by satellite 1, and $w_k$ is AWGN with variance $N=N_0W$. The input symbols $x_k$ must be subject to a peak-power constraint, that can be expressed in the form
\begin{equation}
	|x_k|^2  \leq  P\label{eq:ppc} \,.
\end{equation}

Channel~\eqref{eq:model_pp} under constraint~\eqref{eq:ppc} was completely studied in~\cite{ShBa95}: the capacity-achieving distribution is discrete in amplitude and uniform in phase, and has the following expression
\begin{equation}
	p(r,\theta)=p(\theta)p(r)=\frac{1}{2\pi}\sum_{\ell=1}^m q_\ell \delta(r-p_\ell)\,,\label{eq:distr1}
\end{equation}
with $x_k=r e^{j\theta}$. The distribution is formed of $m$ concentric circles, each having weight $q_\ell$ and radius $p_\ell$. The constraints of the problem, in polar coordinates, become
\begin{eqnarray}
	0 \leq p_\ell \leq \sqrt{P}\label{eq:constraints1}\\
	p_{\ell+1} > p_\ell\\
	0 \leq q_\ell \leq 1\\
	\sum_{\ell=1}^m q_\ell=1\label{eq:constraints4}\,.
\end{eqnarray}

\begin{figure}
	\centering
	\includegraphics[width=1.0\columnwidth]{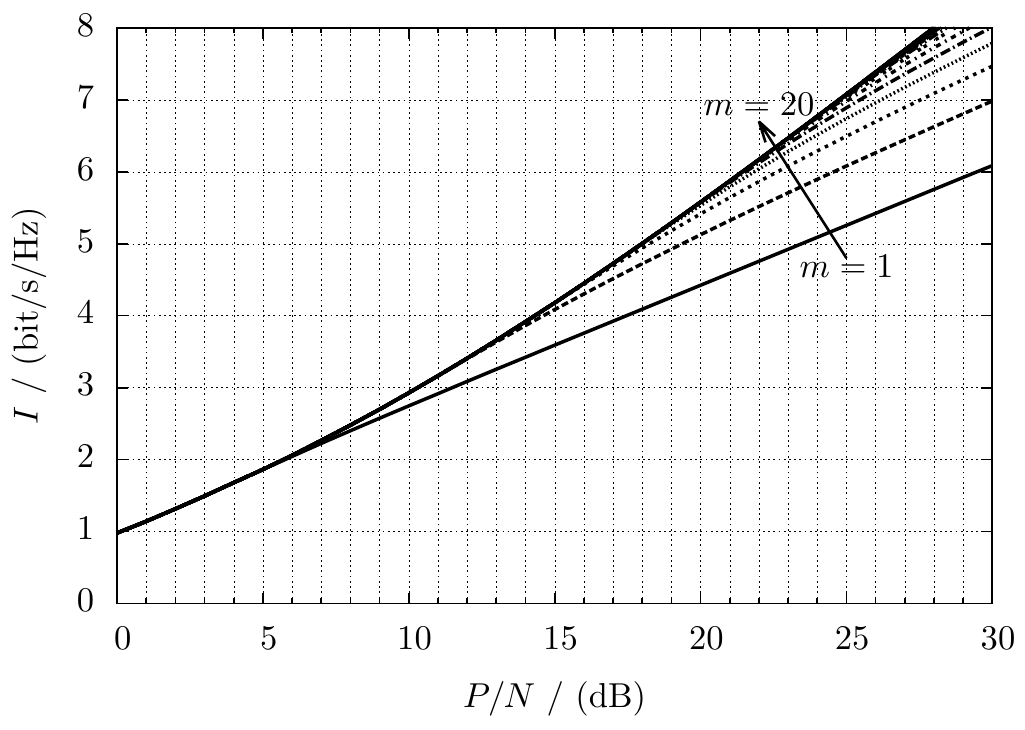}
	\caption{Spectral efficiency for single transmitter for distributions composed of an increasing number of circles $m$ (AWGN channel with peak power constraint).}
	\label{fig:ir1}
\end{figure}

\begin{figure}
	\centering
	\includegraphics[width=1.0\columnwidth]{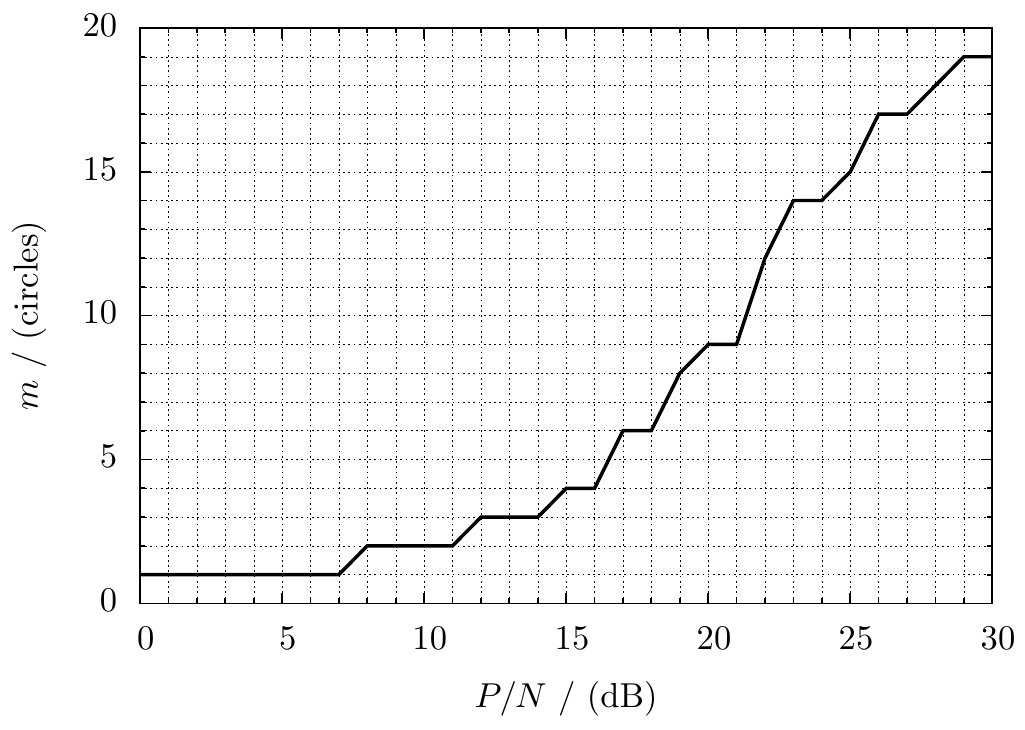}
	\caption{Optimal number of circles of the capacity-achieving distribution for single transmitter.}
	\label{fig:opt_distr}
\end{figure}

For the distribution~\eqref{eq:distr1}, we can compute the rate $I(x_k;y_k)$ in closed form. First of all, we need to derive an expression for the probability density functions (PDFs) of the channel and the observable. Based on the channel model~\eqref{eq:model_pp}, we have
\begin{equation}\label{eq:pyx}
	p(y_k|x_k)=p(y_k|r,\theta)=\frac{1}{\pi N} e^{-\frac{|y_k-r e^{j\theta}|^2}{N}}\,.
\end{equation}
From~\eqref{eq:pyx} we can obtain the PDF $p(y_k)$ as
\begin{eqnarray}
	p(y_k) & = & \int_{r=0}^{+\infty} \int_{\theta=0}^{2\pi}p(y_k|r,\theta)p(r,\theta){\rm d}r{\rm d}\theta\nonumber\\
	& = & \frac{1}{\pi N} \frac{1}{2\pi} \int_{r=0}^{+\infty} \int_{\theta=0}^{2\pi} e^{-\frac{|y_k|^2+|r|^2}{N}} e^{\frac{2\Re\left[y_k r e^{-j\theta}\right]}{N}}\nonumber\\
	& & \cdot \sum_{\ell=1}^m q_\ell \delta(r-p_\ell){\rm d}r{\rm d}\theta\nonumber\\
	& = & \frac{1}{\pi N} \frac{1}{2\pi} \sum_{\ell=1}^m q_\ell e^{-\frac{|y_k|^2+|p_\ell|^2}{N}}\nonumber\\
	& & \cdot \int_{\theta=0}^{2\pi} e^{\frac{2|y_k|p_\ell}{N} \cos(\arg(y_k)-\theta)}{\rm d}\theta\nonumber\\
	& = & \frac{1}{\pi N} \sum_{\ell=1}^m q_\ell e^{-\frac{|y_k|^2+|p_\ell|^2}{N}} \mathrm{I_0}\left(\frac{2|y_k|p_\ell}{N} \right)\,,\label{eq:py}
\end{eqnarray}
where $\mathrm{I_0}(\cdot)$ is the modified Bessel function of the first kind and order zero. By combining~\eqref{eq:pyx} and~\eqref{eq:py} we have
\begin{eqnarray}
	I(x_k;y_k) & = & \mathrm{E}\left[\log_2\frac{p(y_k|x_k)}{p(y_k)}\right]\label{eq:ir1}\\
	& = &\mathrm{E}\left[\log_2{\frac{e^{-\frac{|y_k-r e^{j\theta}|^2}{N}}}{\sum_{\ell=1}^m q_\ell e^{-\frac{|y_k|^2+p_\ell^2}{N}}\mathrm{I_0}\left(\frac{2|y_k|p_\ell}{N} \right)}}\right]\nonumber\,.
\end{eqnarray}
The expectation in~\eqref{eq:ir1} is taken with respect to the actual random variables, i.e., $x_k$ (and hence $r$ and $\theta$) and $y_k$. The latter is a function of $x_k$ and $w_k$, and thus it depends on their statistics. The optimal values of $m$, $q_\ell$ and $p_\ell$ cannot be found in closed form, but they are subject to optimization~\cite{ShBa95}. For this reason, we evaluated~\eqref{eq:ir1} for increasing values of $m$ and, for each value, we optimized the $m$ radii to achieve the highest IR. Optimization results for $1\leq m \leq 20$ are plotted in Fig.~\ref{fig:ir1}. We see that, as expected, as $P/N$ increases, the optimal distribution is formed of a higher number of circles. We also point out that each curve in Fig.~\ref{fig:ir1} is the envelope of all curves with a lower number of circles, so $m$ must be read as the maximum number of circles, i.e., one or more circles can have zero probability. The optimal number of circles is shown in Fig.~\ref{fig:opt_distr} as a function of $P/N$. We point out that the results in Fig.~\ref{fig:ir1} differ from those in~\cite{ShBa95} because of a different SNR definition. In fact, in~\cite{ShBa95}, capacity curves are computed as a function of the SNR per dimension, while our curves are a function of the total SNR.

\subsection{Analysis for Two Transmitters}
Aim of this section is to extend the results of Section~\ref{sec:1tx} to the case of two transmitters. For this scenario, we make the assumption that the optimal distributions of the two inputs are still in the form~\eqref{eq:distr1}. This result has been demonstrated for real inputs~\cite{MaMoKh12}, but not for complex inputs, as the case of interest here. For this reason, the computed IR is a lower bound to the actual channel capacity, whose expression 
is not known. Under this assumption, the input amplitude distributions are
\begin{equation}
	p(r_i)=\sum_{\ell=1}^{m_i} q_{\ell}^{(i)} \delta(r_i-p_{\ell}^{(i)}),\quad i=1,2\nonumber
\end{equation}
and the received signal is an extension of~\eqref{eq:model_pp}:\footnote{We point out that a phase noise term should be considered in the second signal. However, since this shift is assumed to be perfectly known at the receiver and the input distributions are invariant w.r.t. a phase rotation, we do not add it to our model.}
\begin{eqnarray}
	y_k & = & x_k^{(1)}+\gamma x^{(2)}_k+w_k\nonumber\\
	  & = & r_1e^{j\theta_1}+\gamma r_2e^{j\theta_2}+w_k \nonumber\,,
\end{eqnarray}
with each of the two inputs satisfying constraints~\eqref{eq:constraints1}--\eqref{eq:constraints4}. In this scenario, we can express the joint IR $I_{\rm J}=I(x_1,x_2;y)$ as an extension of~\eqref{eq:ir1}:
\begin{equation}\label{eq:ir2}
	I_{\rm J}=\mathrm{E}\left[\log_2{\frac{(2\pi)^2e^{-\frac{|y_k-r_1e^{j\theta_1}-\gamma r_2e^{j\theta_2}|^2}{N}}}{\sum_{\ell=1}^{m_1}\sum_{i=1}^{m_2} q_\ell^{(1)}q_{i}^{(2)}   {\Lambda}_{\ell,i} }}\right]\,,
\end{equation}
where
\begin{equation}
   {\Lambda}_{\ell,i}=\int_0^{2\pi}\int_0^{2\pi} e^{-\frac{|y_k-p_{\ell}^{(1)}e^{j\theta_1}-\gamma p_i^{(2)}e^{j\theta_2}|^2}{N}}\mathrm{d}\theta_1\mathrm{d}\theta_2 \nonumber\,.
\end{equation}

\begin{figure}
	\centering
	\includegraphics[width=1.0\columnwidth]{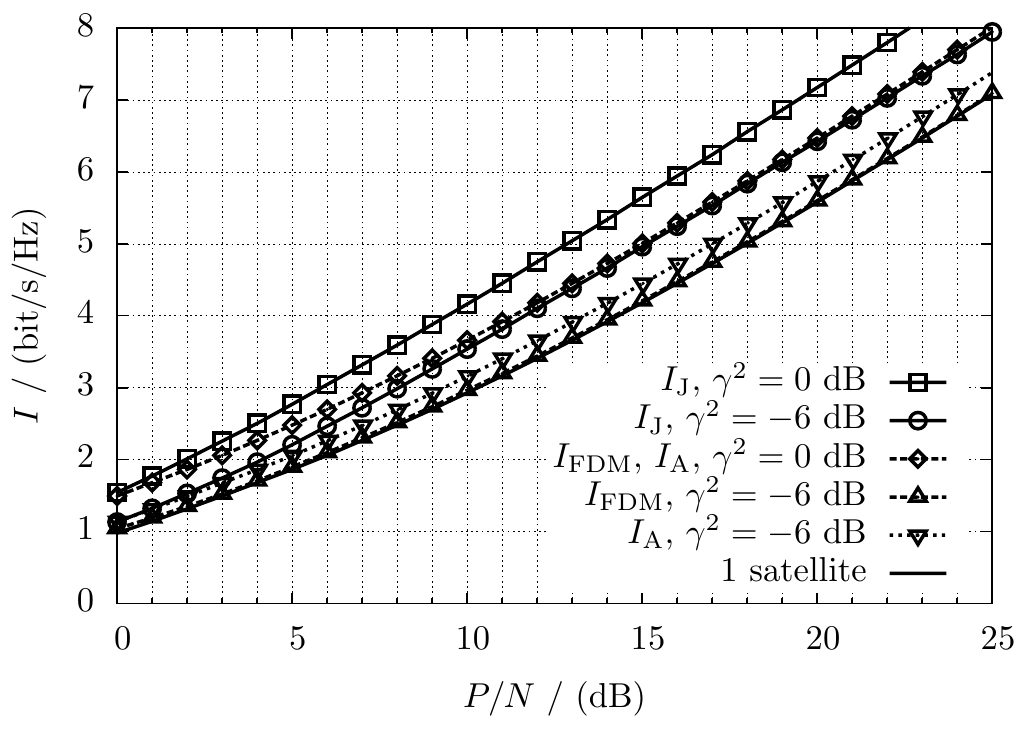}
	\caption{Joint spectral efficiency for different values of $\gamma$ (AWGN channel with peak power constraint).}
	\label{fig:ir_joint}
\end{figure}

We have computed~\eqref{eq:ir2} for different levels of power unbalance between the two received signals; we have verified that the best performance is achieved when using input distributions with only one circle (i.e., $m_1=m_2=1$ in~\eqref{eq:ir2}). The joint IR is shown in Fig.~\ref{fig:ir_joint}, where the curve labeled {\it 1 satellite} is obtained as the envelope of the curves of Fig.~\ref{fig:ir1}. We report here, for comparison, the IR computed when FDM is used, assigning half of the bandwidth to each of the satellites. We see that, unlike the case of average power constraint, FDM is not the optimal choice, not even in the absence of power unbalance (when FDM gains exactly 3 dB from the single satellite). This result comes from a straightforward application of Theorem~\ref{thm:fdm}. In effect, since the two input distributions are not Gaussian, the inequality~\eqref{eq:i1} is strict. Fig.~\ref{fig:ir_joint} also reports the IR $I_{\rm A}$, achievable by the Alamouti scheme. As foreseen by Theorem~\ref{thm:alam}, we see that for $\gamma^2=0$ dB the rate $I_{\rm A}$ is perfectly equivalent to $I_{\rm FDM}$, while for $\gamma^2=-6$ dB FDM performs worse. For all values of $\gamma$ we have that $I_{\rm J}>I_{\rm A}$, since the input signals are not Gaussian processes.

We also point out that, unlike what happens when the constraint is on the average power, in this case the theoretical upper bound for $I_{\rm J}$, when $\gamma=1$, is 6 dB higher than the single transmitter case. This is because, if the two signals are perfectly in phase, the overall signal has double amplitude, and hence its power is $4P$ (i.e., 6 dB higher). This situation is unrealistic (and, in fact, we do not experience a 6 dB gain), 
but it is the upper limit for the IR.

\begin{figure}
	\centering
	\includegraphics[width=1.0\columnwidth]{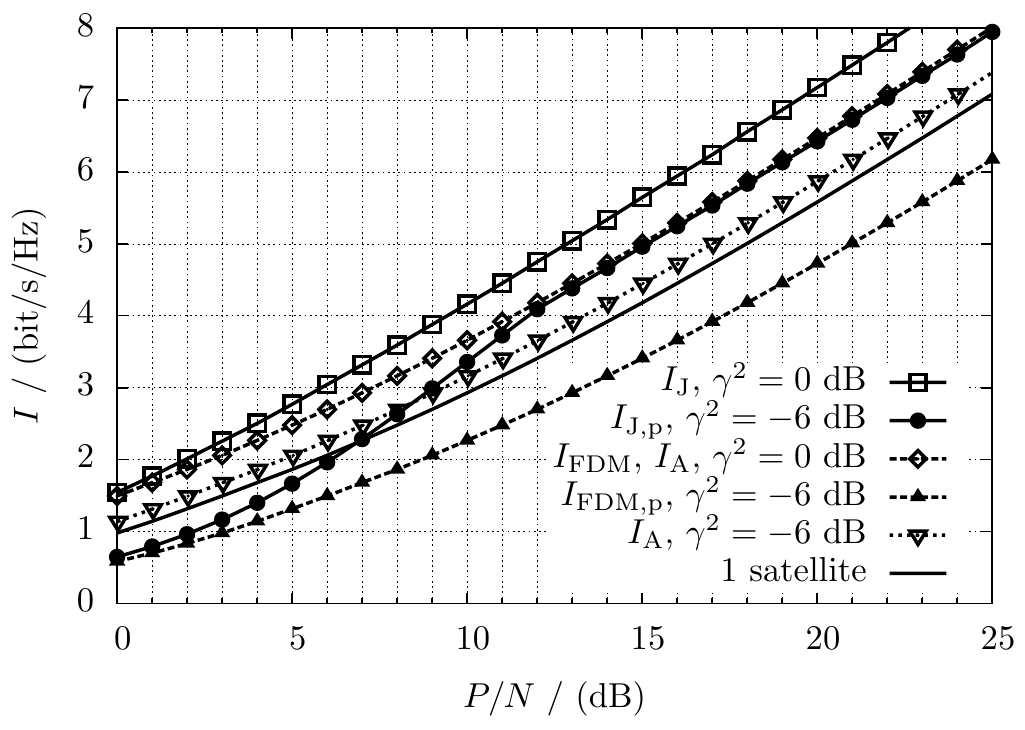}
	\caption{Pragmatic spectral efficiency for different values of $\gamma$ (AWGN channel with peak power constraint).}
	\label{fig:ir_eq}
\end{figure}

As already mentioned, in a broadcast scenario we have the further constraint that the two transmitters must use the same rate. When we impose this constraint to the rates shown in Fig.~\ref{fig:ir_joint}, we obtain the pragmatic rates in Fig.~\ref{fig:ir_eq}. We see again that, as expected, the rates $I_{\rm J,p}$ and $I_{\rm FDM,p}$ have suffered a degradation for $\gamma^2=-6$ dB, and we also notice that, for low values of $P/N$ and a high power unbalance, the use of a single satellite may be convenient over the use of two overlapped signals. However, since the Alamouti scheme is not degraded by the application of the equal rates constraint, we can conclude that the $I_{\rm A}$ grants the best performance in a certain range of $P/N$.

\begin{figure}
	\centering
	\includegraphics[width=1.0\columnwidth]{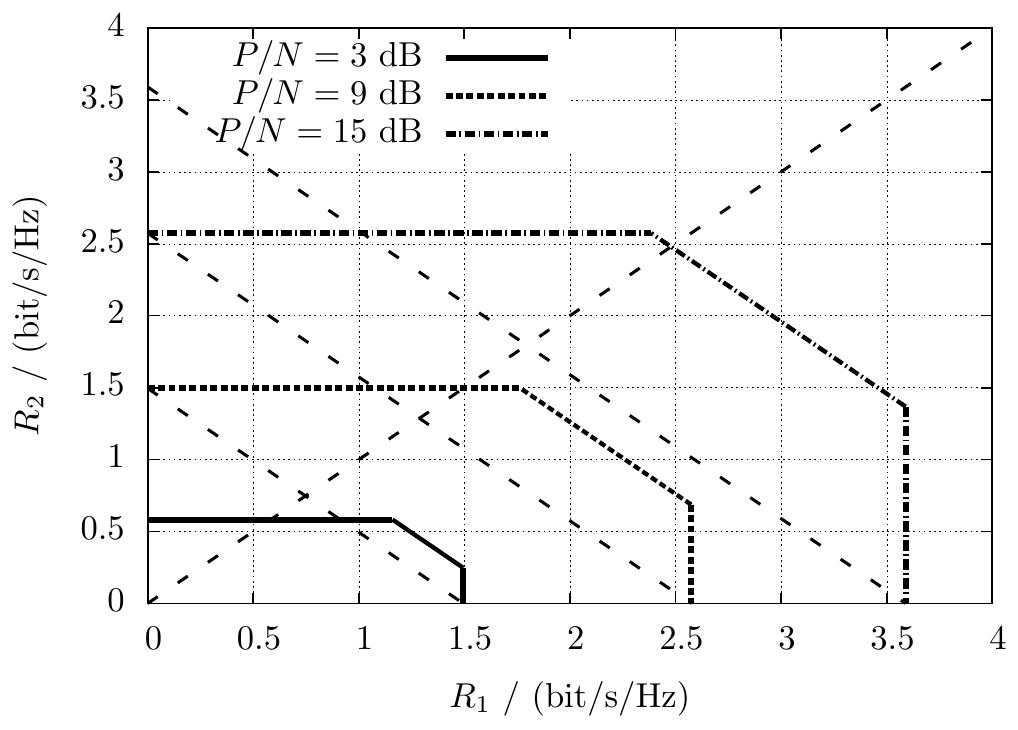}
	\caption{Spectral efficiency regions for $\gamma^2=-6$ dB.}
	\label{fig:cap_reg6}
\end{figure}

We can better understand the impact of the equal rates constraint on the joint IR by studying the SE regions of the channel for different values of $P/N$, reported in Fig.~\ref{fig:cap_reg6}. From the analysis of these figures, we can conclude that the maximum sum-rate cannot always be achieved and we can have a numerical insight of the values of $P/N$ and $\gamma^2$ that allow to improve the rates with respect to a case with only one transmitter. In particular, it is easy to understand that when the power unbalance is low ($\gamma^2\rightarrow0$ dB) the spectral efficiency region is perfectly symmetric and maximum sum-rate is achieved for all values of $P/N$. On the other hand, with high power unbalance ($\gamma^2=-6$ dB) it is clear that maximum sum-rate can be achieved only at high $P/N$, whereas when the power is low the performance of two satellites is worse than that of a single satellite.

\subsection{Practical Constellations for AWGN Channel with Peak Power Constraint}\label{sec:psk_ppc}
We are now interested in evaluating the performance of practical constellations with a finite number of points on the AWGN channel with peak power constraint, in order to find which kind of discrete constellations can be successfully adopted on the satellite channel.

\begin{figure}
	\centering
	\includegraphics[width=1.0\columnwidth]{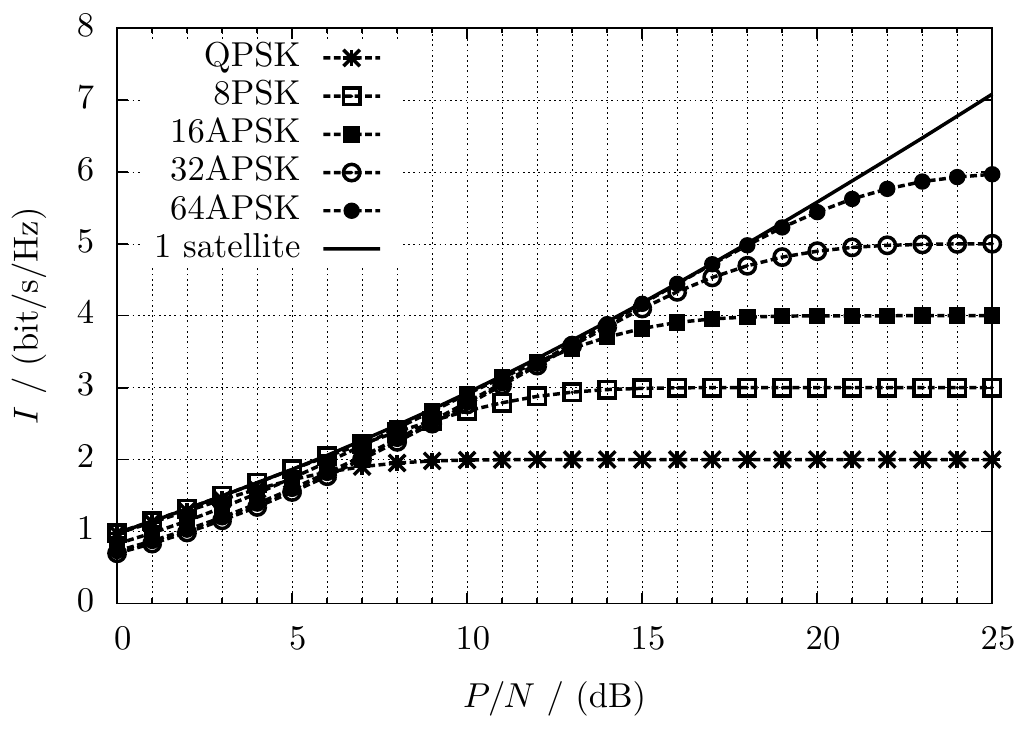}
	\caption{Single transmitter capacity and spectral efficiency for PSK/APSK constellations.}
	\label{fig:ir_apsk_ppc}
\end{figure}

\begin{figure}
	\centering
	\includegraphics[width=1.0\columnwidth]{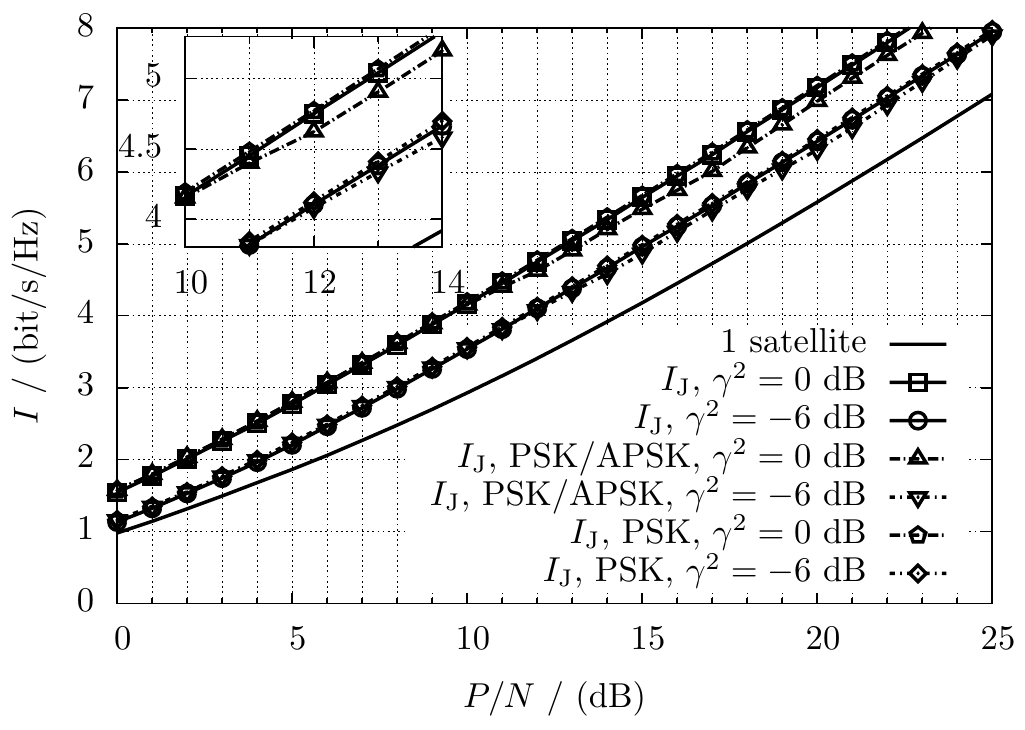}
	\caption{Joint spectral efficiency for PSK/APSK constellations and different values of $\gamma$.}
	\label{fig:ir_ppc2}
\end{figure}

\begin{figure}
	\centering
	\includegraphics[width=1.0\columnwidth]{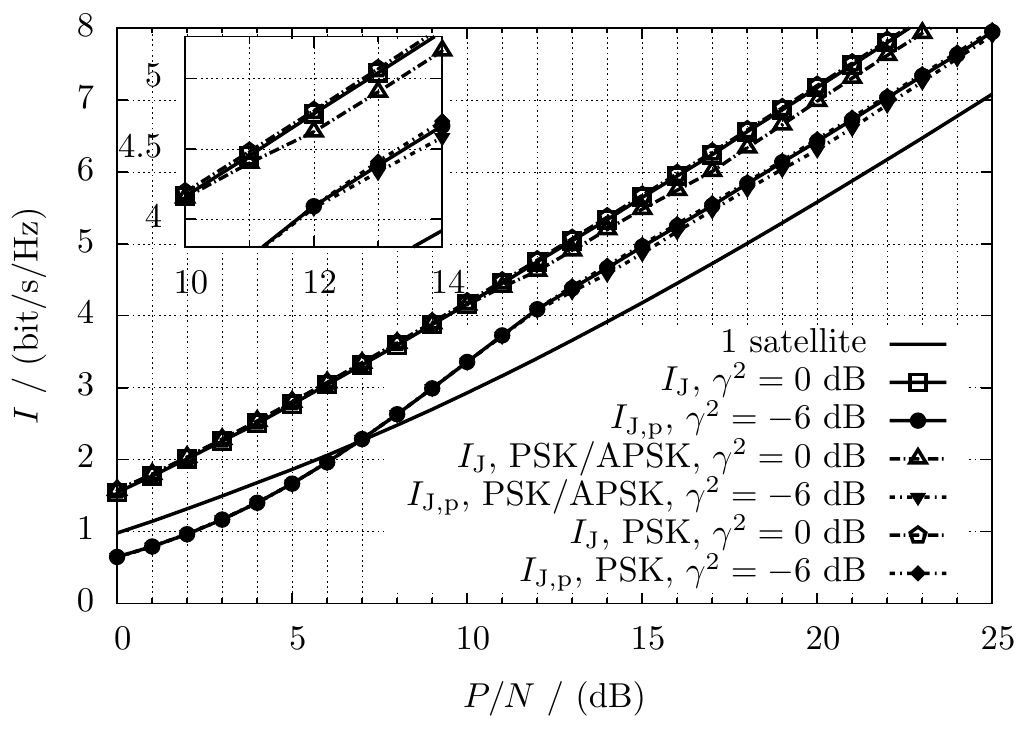}
	\caption{Pragmatic spectral efficiency for PSK/APSK constellations and different values of $\gamma$.}
	\label{fig:ir_ppc2_er}
\end{figure}

Starting from the single transmitter case, we see in Fig.~\ref{fig:ir_apsk_ppc} that $M$-ary phase-shift keying and amplitude-phase-shift keying ($M$PSK/$M$APSK) constellations, usually adopted in satellite communications, are practically capacity-achieving. However, as foreseen also by the theoretical analysis, constellations with multiple circles (such as APSK) are suboptimal when two transmitters are adopted. This can be seen in Fig.~\ref{fig:ir_ppc2}, showing the envelopes of the SEs achievable with quaternary PSK (QPSK), 8PSK, 16APSK, 32APSK and 64APSK, where APSK exhibits a loss with respect to the bound $I_{\rm J}$ for high $P/N$ values. As suggested by the theoretical results, we see that the bound is achieved by replacing APSK constellations with PSKs with the same cardinality, whose envelopes are again shown in the figure. Fig.~\ref{fig:ir_ppc2_er} reports the same analysis for the pragmatic rates, and the same conclusions hold. We point out that FDM and Alamouti schemes perform single-user operations, so they practically achieve their corresponding theoretical bounds with classical PSK/APSK constellations. Finally, we mention that we have attempted an optimization of the constellations, using the same algorithm described in~\cite{UgMoCoPiMiMo14}, imposing that the constellations adopted by the two transmitters are identical. Optimization results suggest that PSKs are practically optimal in this case.

\section{Satellite Channel}\label{sec:sat_ch}

\begin{figure}
	\centering
	\includegraphics[width=1.0\columnwidth]{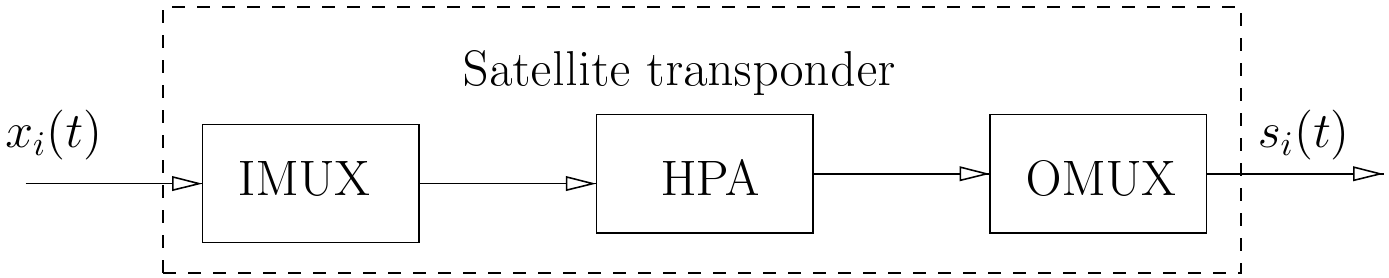}
	\caption{Block diagram of the considered satellite transponder.}\label{fig:bd_satch}
\end{figure}

This section investigates the performance of the system in Fig.~\ref{fig:system} when a realistic satellite transponder model is used. The block diagram of the adopted transponder model is depicted in Fig.~\ref{fig:bd_satch}. It shows an input multiplexing (IMUX) filter which removes the adjacent channels, a HPA, and an output multiplexing (OMUX) filter aimed at reducing the spectral broadening caused by the nonlinear amplifier. The HPA AM/AM and AM/PM characteristics and the IMUX/OMUX impulse responses are those provided in~\cite{DVB-S2-TR}. Particularly, the OMUX filter has -3 dB bandwidth equal to 38 MHz and the HPA has AM/AM and AM/PM characteristics that are called ``conventional'' in~\cite{DVB-S2-TR}. Although the HPA is a nonlinear memoryless device, the overall system has memory due to the presence of IMUX and OMUX filters. 

The transmitted signals at the input of the two satellites are linearly modulated as in~\eqref{eq:signals1}, with the same pulse and symbol interval, and the information symbols $x_k^{(i)}$ are drawn from a discrete constellation. The symbol intervals of the two signals are also assumed to be perfectly aligned. Thus, the received signal reads as in~\eqref{eq:rec}. Process $\phi(t)$ models the difference of phase between oscillators and their phase noise, and is considered perfectly known at the receiver. We employ the adaptive receiver proposed in~\cite{UgMoCoPiMiMo14,DVB-TM-S2}: a sufficient statistic for detection is extracted by using oversampling at the output of a low pass filter~\cite{MeOePo94}, and a fractionally-spaced minimum mean square error (FS-MMSE) equalizer, working at twice the symbol rate, acts as adaptive filter followed by a multiuser detector. The adaptivity is accomplished by means of the least mean square or the recursive least square  algorithms~\cite{Ha96}. The multiuser detector computes the a posteriori probabilities of the symbols as
\begin{equation}
	p\left(y_k|x_k^{(1)},x_k^{(2)}\right) \propto \exp\left\{ -\frac{\left|y_k - \beta \left(x_k^{(1)}+ \gamma x_k^{(2)}e^{j\phi_k} \right) \right|^2}{N_0} \right\}\nonumber
\end{equation}
where $y_k$ is the sample at the output of the FS-MMSE equalizer, $\beta$ is a possible (complex-valued) bias, and $\phi_k=\phi(kT)$ is the phase noise process at the receiver (under the assumption that $\phi(t)$ is slow enough w.r.t. the symbol time). 

\begin{figure}
	\centering
	\includegraphics[width=0.7\columnwidth]{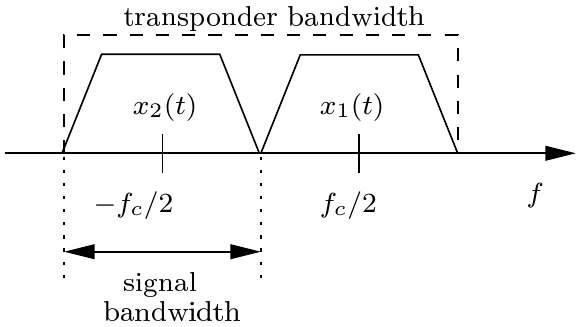}
	\caption{Transponder bandwidth allocation for FDM.}\label{fig:fdm_bw}
\end{figure}

Similarly to previous sections, we also consider an FDM scenario: the transponder bandwidth is equally divided into two subchannels as schematically depicted in Fig.~\ref{fig:fdm_bw}. Then, the FDM receiver performs detection separately with two FS-MMSE equalizers, followed by a symbol-by-symbol receiver.

As already done with the other channel models, we adopt the Alamouti scheme as a third possibility: the Alamouti precoding is performed on transmitted symbols and, at the receiver side, after a proper processing, two separate FS-MMSE equalizers and symbol-by-symbol receivers are adopted. Unlike the previous scenarios, in the presence of nonlinear distortions and phase noise the Alamouti scheme cannot perfectly separate the two signals at the receiver. However we will show in the numerical results that its performance is still excellent.

The complexity of the channel model does not allow (to the best of our knowledge) to obtain results in a closed form as in previous sections. Hence, the achievable SEs for this scenario are computed through the Monte Carlo method proposed in~\cite{ArLoVoKaZe06} (see also~\cite{PiMoCoAl13} for details on the application to the satellite channel). We point out that these values  are a lower bound to the actual SE and they are achievable with the specific adopted receiver. The ensuing SE curves describe the possible achievable gains in a scenario that is more realistic than those described in previous sections. All results will be reported as function of $P_\mathrm{sat}/N$, where $P_{\mathrm{sat}}$ is the HPA power at saturation.

\subsection{Numerical results}
We consider transmitted signals with baudrate 37 Mbaud, adopting the classical constellations of satellite communications, i.e., QPSK, 8PSK, 16APSK and 32APSK (denoted to as PSK/APSK schemes). As an alternative to the classical PSK/APSK, we also consider the use of 16PSK and 32PSK, as suggested by the theoretical analysis. The adopted shaping pulse $p(t)$ has root raised cosine spectrum with roll-off $\alpha=0.1$. The input back-off is set to 0 dB for QPSK and 8PSK, and to 3 dB for all other modulations.\footnote{We found these values to be optimal from other activities beyond this paper. We also point out that the impact of interchannel interference due to transponders transmitting on adjacent frequencies is negligible for all the presented scenarios, and hence it will not be considered~\cite{NGW13}.}

\begin{figure}
	\centering
	\includegraphics[width=1.0\columnwidth]{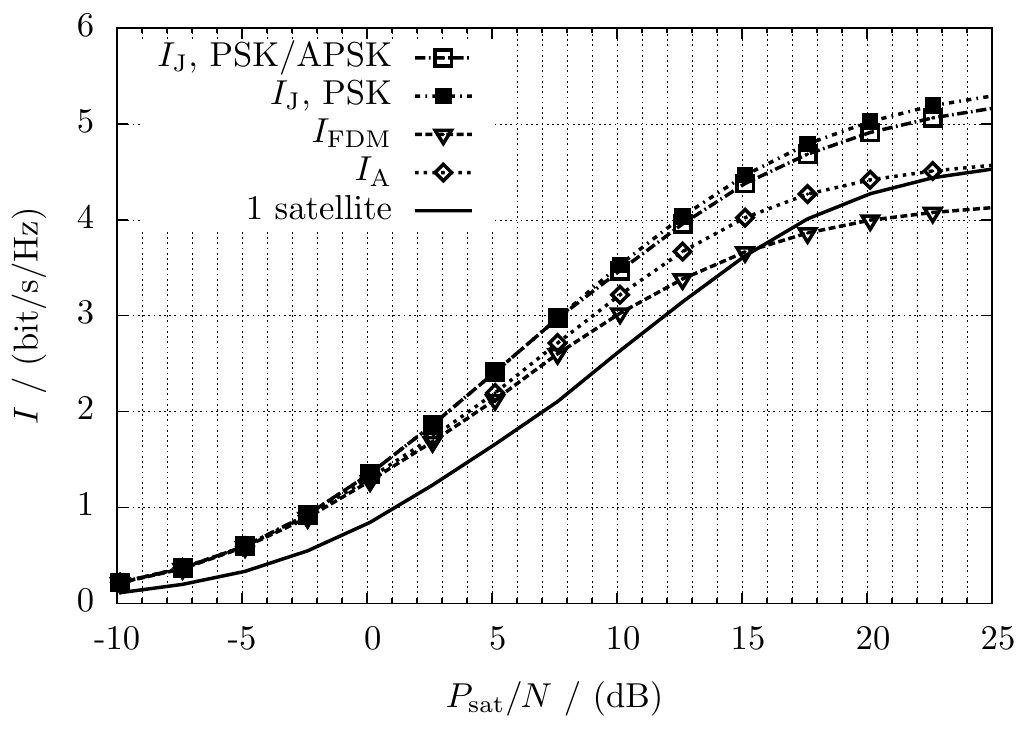}
	\caption{Spectral efficiency achievable by PSK and APSK constellations for two satellites and $\gamma^2=0$ dB.}\label{fig:ase_sat_pun0}
\end{figure}

\begin{figure}
	\centering
	\includegraphics[width=1.0\columnwidth]{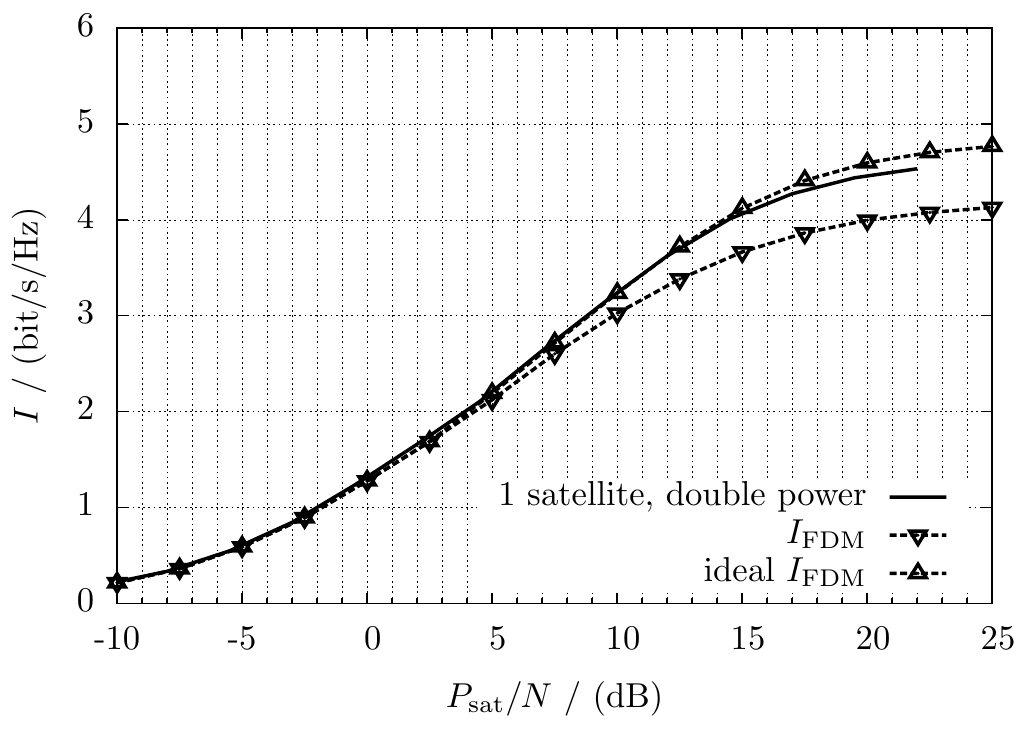}
	\caption{Spectral efficiency achievable by PSK/APSK constellations for two satellites using FDM and $\gamma^2=0$ dB.}\label{fig:ase_sat_fdm_pun0}
\end{figure}

Fig.~\ref{fig:ase_sat_pun0} shows the envelope of the pragmatic SE $I_{\rm J,p}$ for the considered modulations, with power unbalance $\gamma^2=0$ dB. Details on the modulations of the envelope are reported in Table~\ref{tab:env_Pun0}. The figure also shows the SE for FDM, for the Alamouti scheme, and for a single satellite. In case of FDM, each signal has baudrate $1/T=$18.5 Mbaud, and the frequency spacing is equal to $f_c=(1+\alpha)/T$= 20.35~MHz.\footnote{Other values of frequency spacing have been tested, but 20.35 MHz has been found to be practically optimal for this scenario.} We can see from the figure that two overlapped signals can achieve a higher SE than all considered alternatives. Moreover the envelopes show that 32APSK and 32PSK are not convenient in case of overlapped signals, since they perform worse than 16APSK and 16PSK modulations, and PSK modulations perform better than APSK modulations. It is interesting to notice that, although the channel model is affected by nonlinear effects, inequality~\eqref{eq:i3} still holds true even in this case. We also notice that, at high $P_\mathrm{sat}/N$, FDM performs worse even than a single satellite. This loss is due to the interchannel interference (ICI) from the second FDM signal, which lies in the same OMUX bandwidth. In fact, due to the spectral regrowth after the HPA, the two FDM signals are no more orthogonal. This effect is proved in Fig.~\ref{fig:ase_sat_fdm_pun0}, that compares the FDM curve with two SE  curves: ideal FDM in the absence of ICI, and a single satellite with twice the power $P_\mathrm{sat}$. Similarly to the linear channel, ideal FDM can achieve the same SE as the single satellite with double power, but in the actual case ICI has an impact on performance. 

\begin{table}
	\centering
	\begin{tabular}{|c|rcr|}
		\hline
		Modulation & \multicolumn{3}{|c|}{$P_{\mathrm{sat}}/N$ [dB]}  \\
		\hline
		\hline
		QPSK & -10& -- & 0 \\
		\hline 
		8PSK & 0 & -- & 7.5 \\
		\hline 
		16APSK/16PSK & 7.5 & -- & 25 \\
		\hline
	\end{tabular}
	\vspace{2mm}
	\caption{$P_{\mathrm{sat}}/N$ range of the envelope $I_{\mathrm{J,p}}$ for PSK and APSK modulations and $\gamma=1$.}\label{tab:env_Pun0}
\end{table}

\begin{figure}
	\centering
	\includegraphics[width=1.0\columnwidth]{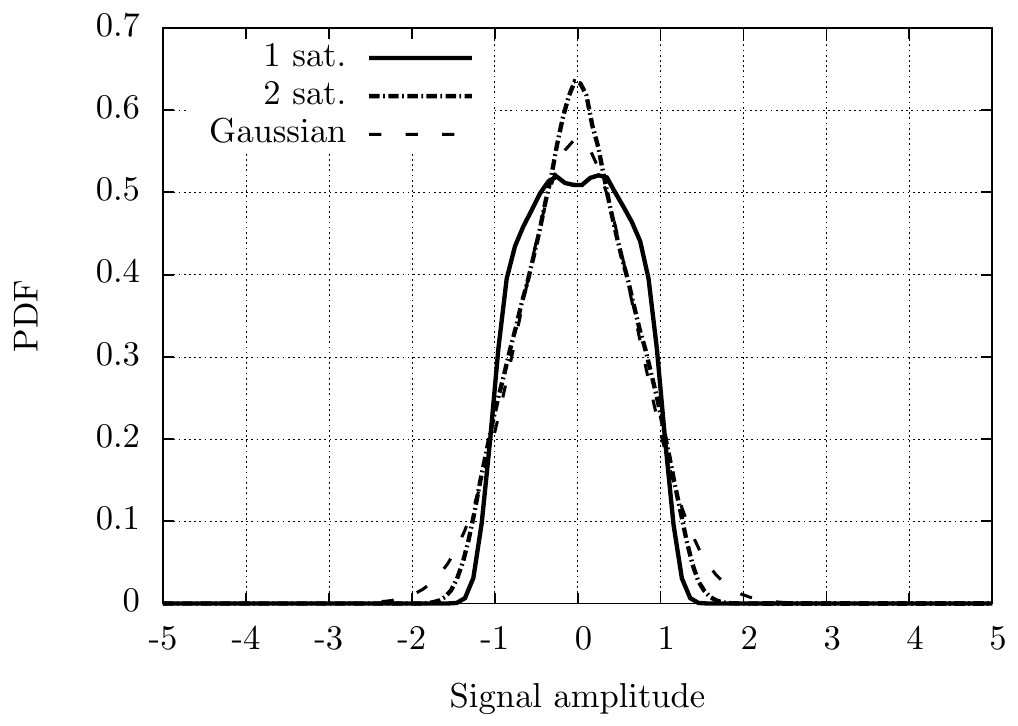}
	\caption{Probability density function of the signal amplitude with 16PSK, RRC \mbox{$\alpha=0.1$}, IBO=3 dB, $\gamma^2=0$ dB.}\label{fig:pdf16}
\end{figure}

\begin{figure}
	\centering
	\includegraphics[width=1.0\columnwidth]{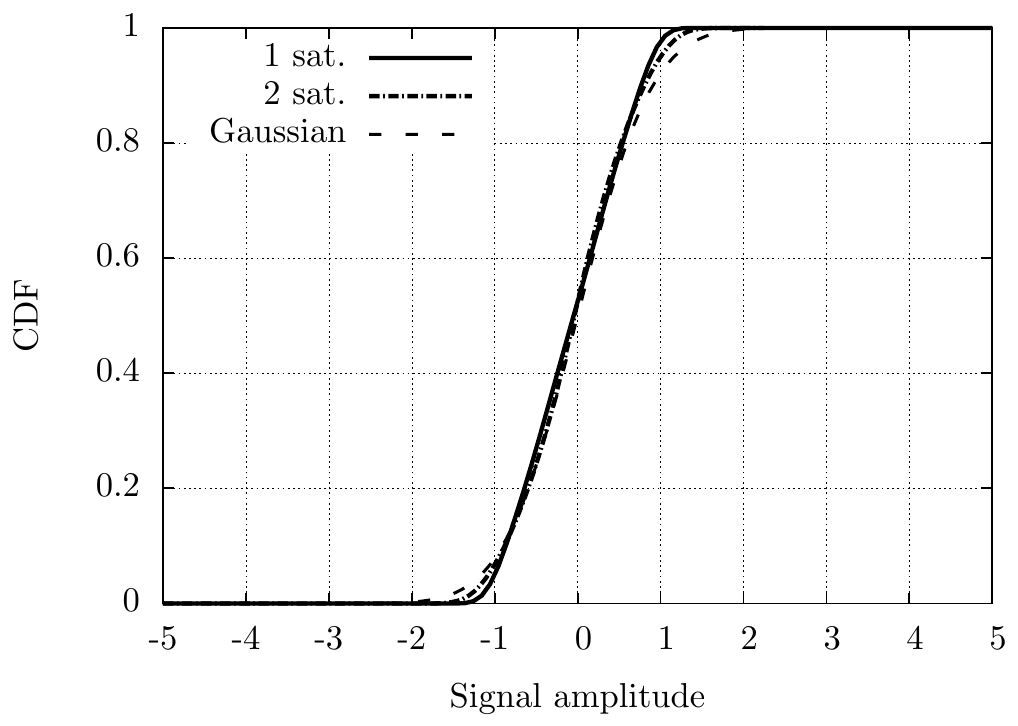}
	\caption{Cumulative distribution function of the signal amplitude with 16PSK, RRC \mbox{$\alpha=0.1$}, IBO=3 dB, $\gamma^2=0$ dB.}\label{fig:cdf16}
\end{figure}

We can notice from Fig.~\ref{fig:ase_sat_pun0} that gains given by two overlapped signals w.r.t. a single satellite can be higher than 3 dB. The gains over 3 dB are related to the shaping of the overall signal, obtained by the sum of the satellite outputs. Indeed, as already mentioned in Section~\ref{sec:fdm}, the sum of two signals has an amplitude distribution that is closer to a Gaussian distribution. Figs.~\ref{fig:pdf16} and~\ref{fig:cdf16} show the PDF and the cumulative distribution function (CDF) of the signal amplitude, properly normalized by the number of transmitting satellites. We compare the amplitude distributions of a single signal and two overlapped signals (with $\gamma^2=0$ dB), when the transmitters adopt 16PSK, RRC pulses with roll-off $\alpha=0.1$, IBO equal to 3 dB. For comparison purpose we report also the PDF and CDF of the Gaussian distribution with unit variance. It is clear from the figures that the sum of two signals is closer to a Gaussian distribution than the single transmitter. We have verified that similar considerations hold for 8PSK.

\begin{figure}
	\centering
	\includegraphics[width=1.0\columnwidth]{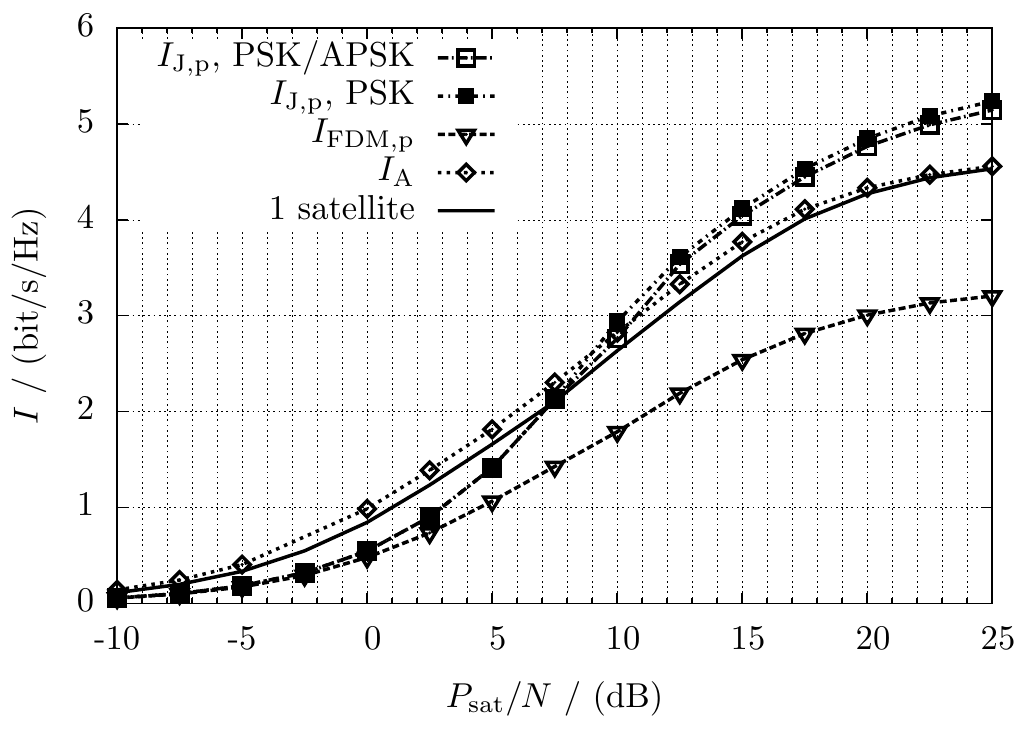}
	\caption{Spectral efficiency achievable by DVB-S2 constellations for two satellites and $\gamma^2=-6$ dB.}\label{fig:ase_sat_pun6}
\end{figure}

\begin{figure}
	\centering
	\includegraphics[width=1.0\columnwidth]{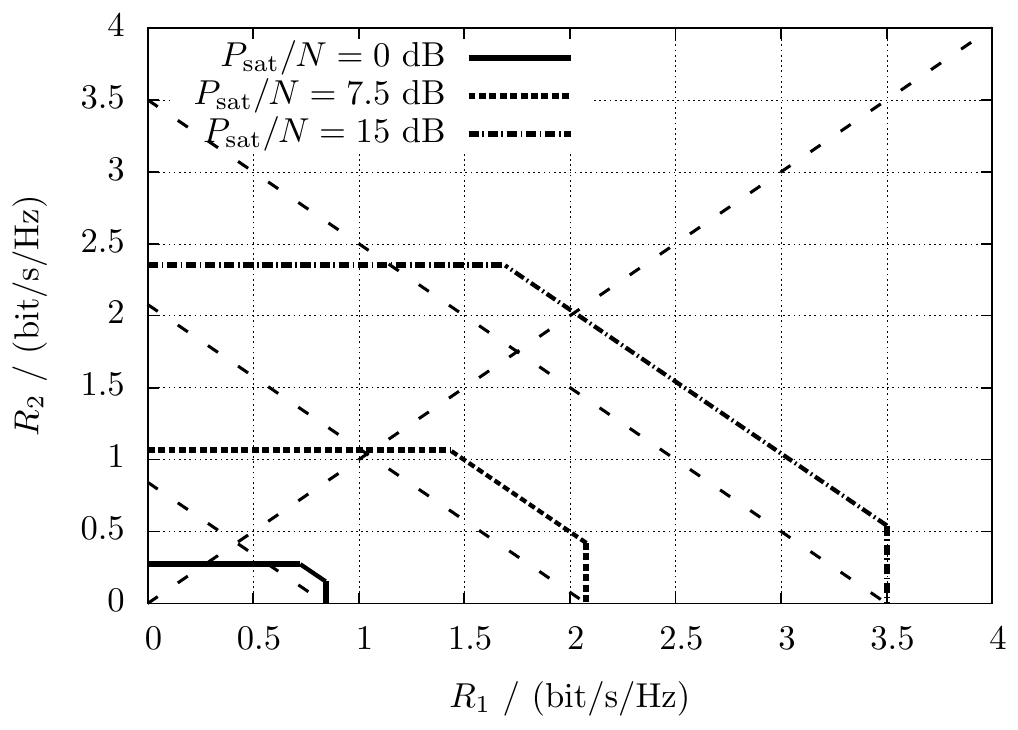}
	\caption{Spectral efficiency regions for DVB-S2 constellations for two satellites and $\gamma^2=-6$ dB.}\label{fig:ase_sat_pun6_region}
\end{figure}

Fig.~\ref{fig:ase_sat_pun6} and Table~\ref{tab:env_Pun6} report SE curves for the same scenario as in Fig.~\ref{fig:ase_sat_pun0}, but with power unbalance equal to 6 dB. Overlapped signals again outperform FDM for every $P_\mathrm{sat}/N$ value but, since the equal rate constraint limits the performance to that of the lower power signal, the Alamouti scheme and a single satellite have higher SE at low $P_\mathrm{sat}/N$. The behavior of $I_{\rm J,p}$ w.r.t. a single satellite can be seen from the SE regions in Fig.~\ref{fig:ase_sat_pun6_region}, and it can be noticed that it is perfectly in line with results found for the peak limited AWGN channel, despite a huge difference between the two models.

\begin{table}
	\centering
	\begin{tabular}{|c|rcr|}
		\hline
		Modulation & \multicolumn{3}{|c|}{$P_{\mathrm{sat}}/N$ [dB]}  \\
		\hline
		\hline
		QPSK & -10& -- & 5 \\
		\hline 
		8PSK & 5 & -- & 12.5 \\
		\hline 
		16APSK/16PSK & 12.5 & -- & 25 \\
		\hline
	\end{tabular}
	\vspace{2mm}
	\caption{$P_{\mathrm{sat}}/N$ range of the envelope $I_{\mathrm{J,p}}$ for PSK and APSK modulations and $\gamma=1/2$.}\label{tab:env_Pun6}
\end{table}

\section{Conclusions}\label{sec:conclusion}
We investigated  the rates achievable by a system using two co-located satellites. We exploited the second satellite to improve the spectral efficiency. We studied three models: AWGN channel with average power constraint, AWGN channel with peak power constraint, and the DVB-S2 satellite channel. For all cases we considered signals with overlapping frequencies, FDM, and the Alamouti scheme. Overlapped signals resulted to be convenient in all cases w.r.t. FDM, but we showed that there are cases in which the Alamouti scheme can outperform both, and that even a single satellite can be convenient over overlapped signals: these cases depend on the power unbalance and on the received signal-to-noise power ratio.

\appendices

\section{Alamouti Scheme with Time Misalignment} \label{app:alamouti}
In this appendix, we show an alternative implementation of the Alamouti scheme, for the case when the signals received by the two satellites have a time misalignment. The precoding and decoding schemes adopt complex conjugation and time reversal of the signals. A similar precoding has been proposed in~\cite{LiXi07} for orthogonal frequency division multiplexing schemes with two relay nodes.

Let us consider a channel for which, when $x_1(t)$ and $x_2(t)$ of finite duration are transmitted, the received signal is
$$
	y_{a}(t)= x_1(t) + \gamma e^{j\phi}x_2(t-\tau)+ w_{a}(t)
$$
where $\tau$ denotes the time misalignment. After the transmission of these signals, the transmitter sends $x_2^*(-t)$ and $-x_1^*(-t)$. The received signal, in this case, is
$$
	y_{b}(t)= x_2^*(-t) - \gamma e^{j\phi}x_1^*(-t+\tau)+ w_{b}(t)\,.
$$
If the receiver has perfect knowledge of $\tau$, $\phi$, and $\gamma$, it can elaborate the received signals as
\begin{eqnarray}
	\frac{y_{a}(t)-\gamma e^{j\phi}y_{b}^*(-t+\tau)}{\sqrt{1+\gamma^2}} & = & \sqrt{1+ \gamma^2}x_1(t) + \tilde{w}_{a}(t) \nonumber \\
	\frac{y_{b}^*(-t)+\gamma e^{-j\phi}y_{a}(t+\tau)}{\sqrt{1+\gamma^2}} & = & \sqrt{1+ \gamma^2}x_2(t) + \tilde{w}_{b}(t) \nonumber  \,,
\end{eqnarray}
where the Gaussian processes $\tilde{w}_{a}(t)$ and $\tilde{w}_{b}(t)$ are statistically equivalent to  $w_a(t)$ and $w_b(t)$. 
Hence, signals $x_1(t)$ and $x_2(t)$ can be independently detected.

In the presence of phase noise (i.e., when the phase $\phi$ is not constant) and nonlinear effects, however, this scheme works only approximately. The loss due to residual interference will be negligible if signals $x_1(t)$ and $x_2(t)$ have a duration shorter than the phase noise coherence time and nonlinear effects are limited. In particular, the AM/PM characteristic of the nonlinear amplifier must be such that when conjugating the input, the output results to be conjugated too.

\section{Proof of Theorem~\ref{thm:fdm}} \label{app:fdm}
We now prove Theorem~\ref{thm:fdm}; we first prove a preliminary result concerning the differential entropies of two continuous random variables.
\begin{lemma}\label{lemma:l1}
	Let $x$ and $y$ be two independent continuous complex random variables, with probability density functions $p(x)$ and $p(y)$ and differential entropies $h(x)$ and $h(y)$. Then
	\begin{equation}
		h(x+y) \geq 1 + \frac{h(x)+h(y)}{2}  \label{eq:lemma}
	\end{equation}
	with equality if and only if $x$ and $y$ are independent Gaussian random variables with the same variance.
\end{lemma}
\begin{IEEEproof}
	For the entropy power inequality
	\begin{eqnarray}
		2^{h(x+y)} & \geq & 2^{h(x)} + 2^{h(y)} \label{eq:step1} \\ 
				& = & 2^{1+\frac{h(x)+h(y)}{2}} \cosh\left( \frac{h(x)-h(y)}{2} \ln 2\right)\nonumber \\
				& \geq & 2^{1+\frac{h(x)+h(y)}{2}} \label{eq:step3}
	\end{eqnarray}
	where equalities in~\eqref{eq:step1} and~\eqref{eq:step3} hold if and only if $x$ and $y$ are Gaussian and have same variance.  Eq.~\eqref{eq:lemma} is finally derived by taking the logarithm of~\eqref{eq:step3}.
\end{IEEEproof}

We then consider the rates achievable by FDM. Under the assumption of ideal FDM transmission, a sufficient statistic is obtained by sampling the continuous waveforms. The observables for the two subchannels are
\begin{eqnarray}
	y_1 & = & x_1+w_1 \label{eq:fdm1_thm} \\
	y_2 & = & \gamma x_2+w_2  \label{eq:fdm2_thm}
\end{eqnarray}
where $x_1$ and $x_2$ are the signal samples, $w_1$ and $w_2$ are white Gaussian noise processes with power $N/2$ instead of $N$, since FDM works with half the bandwidth w.r.t. the case of a single transmitter. The mutual information of FDM is the average of the mutual information for the two channels, i.e.,
\begin{equation}
I_{\rm FDM} = \frac{h(y_1)+h(y_2)}{2}- \log_2\left( \pi e \frac{N}{2}\right) \,, \nonumber \\ 
\end{equation}
and the pragmatic rate is
\begin{equation}
	I_{\rm FDM,p}= h(y_2) - \log_2\left( \pi e \frac{N}{2}\right) \,. \nonumber
\end{equation}
Since the mutual information is a non decreasing function of the SNR \cite{GuShVe05}, clearly it is $I_{\rm FDM,p}\leq I_{\rm FDM}$.

We finally prove Theorem~\ref{thm:fdm}.
\begin{IEEEproof}
	We first prove inequalities~\eqref{eq:i1} and 
	\begin{equation}
		2I_{\mathrm{2}} \geq I_{\rm FDM,p}\,.\label{eq:i2}
	\end{equation}
	The samples at the output of channel~\eqref{eq:chJ} can be equivalently expressed as $y=y_1+y_2$ and the mutual information of this equivalent expression reads $I_{\mathrm{J}}=h(y_1+y_2)- \log_2(\pi e N)$. Hence,
	\begin{equation}
		I_{\mathrm{J}}-I_{\rm FDM} = h(y_1+y_2)- \frac{h(y_1)+h(y_2)}{2} - 1  \nonumber
	\end{equation}
	from which, by an application of the Lemma, we derive inequality \eqref{eq:i1}.
	The mutual information $I_2$, instead, reads
	\begin{eqnarray}
		I_2 & = & h(y|x_1) - \log_2(\pi e N)\nonumber \\
		& = & h(y_2+w_1) - \log_2(\pi e N )\nonumber
	\end{eqnarray}
	and
	$$
		2I_2- I_{\rm FDM,p} = 2h(y_2+w_1) - h(y_2) - \log_2( 2 \pi e N ) \, 
	$$
	which becomes~\eqref{eq:i2} from Lemma.
	
	Since $I_{\mathrm{J,p}}=\min \left(I_\mathrm{J}, 2I_2 \right)$ and $I_{\rm FDM} \geq I_{\rm FDM,p}$, clearly~\eqref{eq:i3} follows with equality if and only if $x_1$ and $x_2$ are Gaussian with $\gamma^2=0$ dB.
\end{IEEEproof}

\section{Proof of Theorem~\ref{thm:alam}} \label{app:alam}
We now prove Theorem~\ref{thm:alam}.
\begin{IEEEproof}
		Let us start by first proving inequality $(a)$. The observable for Alamouti precoding is
		
	\begin{eqnarray*}
		y_{{\rm A},1} & = & x_1 + \gamma x_2 + w_{{\rm A},1} \\
		y_{{\rm A},2}^* & = & x_2 - \gamma x_1 + w_{{\rm A},2} 
	\end{eqnarray*}
	where $w_{{\rm A},1},w_{{\rm A},2}$ are independent Gaussian random variables with power $N$.
	
	Let us evaluate
	\begin{eqnarray}
		I_{\rm J}-I_{\rm A} & = & h(y) - \frac{h(y_{{\rm A},1},y^*_{{\rm A},2})}{2}  \\
		& \geq & h(y) - \frac{ h(y_{{\rm A},1}) + h(y^*_{{\rm A},2}) }{2} \label{eq:indip}\\
		& = & 0 \label{eq:zero}
	\end{eqnarray}			
	where 	\eqref{eq:zero} is obtained by observing that $h(y)=h(y_{{\rm A},1})=h(y^*_{{\rm A},2})$.
	Equality in~\eqref{eq:indip} is achieved if and only if $y_{{\rm A},1}$ and $y^*_{{\rm A},2}$ are independent. From an application of Lukacs-King  theorem \cite{LuKi54}, independence holds if and only if $x_1$ and $x_2$ are independent Gaussian random variables with the same power.
	
	We now prove inequality $(b)$: the Alamouti observable, after receiver processing, is
	\begin{equation}
		\tilde{y}_{{\rm A},i}= \sqrt{1+\gamma^2}x_i + \tilde{w}_{{\rm A},i} \quad i=1,2 \,. \label{eq:al_eq}
	\end{equation}	 	
	and is still a sufficient statistic for detection. The SNR in \eqref{eq:al_eq} is $(1+\gamma^2)/2 \leq 1$ times the one in \eqref{eq:fdm1_thm} and $(1+\gamma^2)/2\gamma^2\geq 1$ times that in \eqref{eq:fdm2_thm}. Hence, since the mutual information is a concave function of the SNR \cite{GuShVe05}, no matter the input distribution, inequality $(b)$ is straightforward and it holds with equality if and only if $\gamma^2=0$ dB. 
\end{IEEEproof}

\section{Ergodic Rates for Satellites Transmitting the Same Signal}
\label{app:ergodic}
We demonstrate that, when the two satellites transmit  the same signal, the information rate cannot be higher than the case when transmitting independent signals.

\begin{IEEEproof}
When the same signal is transmitted, we obtain a time-varying frequency selective channel as in \eqref{eq:fs_ch}. The received signal can be expressed, by using a matrix notation, as
\begin{equation}
\by=\bH\bx+\bw \label{eq:matrix} \,,
\end{equation}
where $\bx=[x_0,\dots,x_{n-1}]^T$ is the vector of transmitted symbols, $\bw$ is the noise vector, and $\bH$ is a matrix with elements $\bH_{k,i}=h_{k,i}$. In this section we denote the Hermitian operator by $\dagger$, and the identity matrix by $\boldsymbol{I}$.

Matrix $\bH$, for the channel we are considering, can be written in the following form
$$
	\bH= \boldsymbol{I} + \gamma \boldsymbol{\Phi}\boldsymbol{\tilde{H}} \,,
$$
where matrix $\boldsymbol{\Phi}$ is diagonal with elements $\boldsymbol{\Phi}_{k,k}=e^{j\phi_k}$  and $\boldsymbol{\tilde{H}}$ is Toeplitz with elements $\boldsymbol{\tilde{H}}_{k,i}= \mathrm{sinc}((i-k)T-\tau)$.

We denote the singular value decomposition of $\bH$ as $$\bH= \boldsymbol{U}\bL\boldsymbol{V}^{\dagger}\,,$$ where $\boldsymbol{U}, \boldsymbol{V}^{\dagger}$ are unitary matrices, and $\bL$ is a diagonal matrix with elements $\bL_{k,k}=\sigma_k$.

If we set $\tilde\bx=\boldsymbol{V}^{\dagger}x$ and $\tilde{\by}=\boldsymbol{U}^\dagger\by$,  channel \eqref{eq:matrix} becomes $n$ parallel channels in the form
\begin{equation}
\tilde{y}_k= \sigma_k \tilde{x}_k + \tilde{w}_k \,,  \label{eq:sch}
\end{equation}	
with SNR $\sigma_k^2P/N$.
Say $I(\sigma_k^2P/N)$ the mutual information of~\eqref{eq:sch} as function of the SNR, the information rate is
\begin{eqnarray}
	\frac{I(\bx;\by|\bH)}{n} & = &  \frac{1}{n}\sum_{k=0}^{n-1} I\left( \sigma_k^2 \frac{P}{N} \right) \nonumber  \\
	& \leq &  I\left(  \frac{1}{n}\sum_{i=0}^{n-1} \sigma_k^2 \frac{P}{N} \right) \nonumber \\
	& \leq &  I\left(  \frac{1}{n} \mathrm{trace}\left( \bH^{\dagger}\bH\right) \frac{P}{N} \right)  \,,\nonumber 
\end{eqnarray}
where the last inequality is due to the concavity of the mutual information as a function of the SNR \cite{GuShVe05}. The trace can be rewritten as
$$
\mathrm{trace}\left( \bH^{\dagger}\bH\right) = \mathrm{trace}\left( \boldsymbol{I} + 2\gamma\mathcal{R}(\boldsymbol{\Phi}\boldsymbol{\tilde{H}})   + \gamma^2 \boldsymbol{\tilde{H}^{\dagger}}\boldsymbol{\tilde{H}}\right) 
\,.$$
Since we are dealing with an ergodic process, it holds that
\begin{eqnarray}
	\lim_{n\rightarrow \infty} \frac{1}{n} \mathrm{trace}\left( \bH^{\dagger}\bH\right) & = & 
	\lim_{n\rightarrow \infty} \frac{1}{n} \mathrm{trace}\left( \boldsymbol{I} + \gamma^2 \boldsymbol{\tilde{H}^{\dagger}}\boldsymbol{\tilde{H}}  \right) \nonumber \\
	& = & 1+\gamma^2 \nonumber
\end{eqnarray}
where last equality is found through the Szeg{\"o} Theorem \cite{Gr06}.
We finally obtain the following inequality
\begin{eqnarray}
	\lim_{n\rightarrow \infty} \frac{I(\bx;\by|\bH)}{n} & \leq &  I\left(  (1+\gamma^2)\frac{P}{N} \right) \, . \nonumber
\end{eqnarray}
The right term is independent of $\boldsymbol{U}$ and $ \boldsymbol{V}^{\dagger}$. Hence, two satellites transmitting the same signal  have achievable information rate lower than the one achievable with the Alamouti scheme and same input distribution.
\end{IEEEproof}
\bibliographystyle{ieeetr}

\end{document}